\documentclass[journal]{IEEEtran}

\ifCLASSINFOpdf
   \usepackage[pdftex]{graphicx}
\else
\fi

\usepackage{comment}

\usepackage{amsmath}
\usepackage{mathtools}

\usepackage{acronym}

\usepackage{amsfonts}
\usepackage{amsthm}
\usepackage{thmtools}

\declaretheorem[name=Theorem]{theorem}

\usepackage{siunitx}
\usepackage{stfloats}
\usepackage{mleftright}
\usepackage{bm}
\usepackage{bbm}
\usepackage{enumitem}
\usepackage{multirow}

\usepackage{placeins}
\usepackage{tikz}
\usepackage{pgfplots}
\usepackage{scalerel}
\usepgfplotslibrary{groupplots}

\usepackage[textsize=tiny]{todonotes} %

\usetikzlibrary{external}
\definecolor{tab101}{RGB}{228,26,28}
\definecolor{tab102}{RGB}{55,126,184}
\definecolor{tab103}{RGB}{77,175,74}
\definecolor{tab104}{RGB}{152,78,163}
\definecolor{tab105}{RGB}{255,127,0}

\definecolor{cR1}{rgb}{0,.59,.51}  %
\definecolor{cR2}{rgb}{.27,.39,.66} %
\definecolor{cR3}{rgb}{.63,.13,.13} %

\definecolor{cR4}{rgb}{.64,.06,.48} %
\definecolor{cR5}{rgb}{.87,.60,.10} %
\definecolor{cR6}{rgb}{.14,.63,.87} %

\definecolor{cR7}{RGB}{100, 221, 23} %
\definecolor{cR8}{RGB}{167,130,46} %
\definecolor{cR9}{rgb}{0.4, 0.6, 0.2} %

\ifCLASSOPTIONcompsoc
  \usepackage[caption=false,font=normalsize,labelfont=sf,textfont=sf]{subfig}
\else
  \usepackage[caption=false,font=footnotesize]{subfig}
\fi

\usepackage{graphicx}
\usepackage{color}
\usepackage{booktabs}
\usepackage{pgf, tikz, pgfplots}
\usepgfplotslibrary{fillbetween}
\usetikzlibrary{shapes,arrows.meta, calc}

\usepackage[ruled,vlined, linesnumbered]{algorithm2e}
\let\oldnl\nl%
\newcommand{\nonl}{\renewcommand{\nl}{\let\nl\oldnl}}%

\SetCommentSty{mycommfont}

\newcommand\hIl{h_{\text{I},\ell}}
\newcommand\hRl{h_{\text{R},\ell}}

\definecolor{KITgreen}{rgb}{0,.59,.51}
\definecolor{KITpalegreen}{RGB}{130,190,60} 
\definecolor{KITblack}{rgb}{0,0,0}
\definecolor{KITblue}{rgb}{.27,.39,.66}
\definecolor{KITred}{rgb}{.63,.13,.13}
\definecolor{KITpurple}{rgb}{.64,.06,.48}
\definecolor{KITcyan}{rgb}{.14,.63,.87}
\definecolor{KITyellow}{rgb}{.98,.89,0}
\definecolor{KITorange}{rgb}{.87,.60,.10}

\definecolor{EMBPcolor}{rgb}{0,.59,.51}
\colorlet{BPcolor}{magenta}
\definecolor{EMNBPcolor}{RGB}{130,190,60} 
\definecolor{EMBPLCcolor}{rgb}{.63,.13,.13}
\definecolor{EMBPDiraccolor}{rgb}{.14,.63,.87}
\definecolor{VAEcolor}{rgb}{.87,.60,.10}
\colorlet{RNNcolor}{KITblue}
\colorlet{ViterbiNetcolor}{KITcyan}
\colorlet{decdircolor}{KITred}

\definecolor{cR1}{rgb}{0,.59,.51}  %
\definecolor{cR2}{rgb}{.27,.39,.66} %

    \acrodef{APP}[APP]{a posteriori probabilitie}
    \acrodef{AWGN}[AWGN]{additive white Gaussian noise}
    \acrodef{BP}[BP]{belief propagation}
    \acrodef{BPSK}[BPSK]{binary phase-shift keying}
    \acrodef{BER}[BER]{bit error rate}
    \acrodef{BMD}[BMD]{bit-metric decoder}
    \acrodef{BMI}[BMI]{bitwise mutual information}
    \acrodef{BICM}[BICM]{bit-interleaved coded modulation}
    \acrodef{CMA}[CMA]{constant modulus algorithm}
    \acrodef{CSI}[CSI]{channel state information}
    \acrodef{DD}[DD]{decision-directed}
    \acrodef{DNN}[DNN]{deep neural network}
    \acrodef{ELBO}[ELBO]{evidence lower bound}
    \acrodef{EM}[EM]{expectation maximization}
    \acrodef{FEC}[FEC]{forward error correction}
    \acrodef{FIR}[FIR]{finite impulse response}
    \acrodef{iid}[i.i.d.]{independent and identically distributed}
    \acrodef{ISI}[ISI]{inter-symbol interference}
    \acrodef{KL}[KL]{Kullback-Leibler}
    \acrodef{LDPC}[LDCP]{low-density parity-check}
    \acrodef{LLR}[LLR]{log-likelihood ratio}
    \acrodef{MAP}[MAP]{maximum a posteriori}
    \acrodef{ML}[ML]{maximum likelihood}
    \acrodef{MMSE}[LMMSE]{linear minimum mean squared error}
    \acrodef{MSE}[MSE]{mean squared error}
    \acrodef{NBP}[NBP]{neural belief propagation}
    \acrodef{NN}[NN]{neural network}
    \acrodef{pdf}[PDF]{probability density function}
    \acrodef{pdp}[PDP]{power-delay profile}
    \acrodef{pmf}[PMF]{probability mass function}
    \acrodef{QPSK}[QPSK]{quadrature phase-shift keying}
    \acrodef{RNN}[RNN]{recurrent neural network}
    \acrodef{SNR}[SNR]{signal-to-noise ratio}
    \acrodef{SPA}[SPA]{sum-product algorithm}
    \acrodef{VAE}[VAE]{variational autoencoder}
    \acrodef{VAELE}[VAE-LE]{variational autoencoder-based linear equalizer}

\newcommand{\Estep}{\mbox{E-step}}
\newcommand{\Mstep}{\mbox{M-step}}

\pgfplotsset{compat=newest}

\begin{document}
\title{Blind Channel Estimation and Joint Symbol Detection with Data-Driven Factor Graphs}

\author{Luca Schmid, Tomer Raviv, Nir Shlezinger,~\IEEEmembership{Senior~Member,~IEEE}, and Laurent Schmalen,~\IEEEmembership{Fellow, IEEE}%
\thanks{This work has received funding in part from the European Research Council (ERC) under the European Union’s Horizon 2020 research and innovation programme (grant agreement No. 101001899), in part from the German
Federal Ministry of Education and Research (BMBF) within the project Open6GHub (grant agreement 16KISK010), and in part from the Israeli Ministry of Science and Technology.}%
\thanks{This work will be presented at the Asilomar Conference on Signals, Systems, and Computers 2024, Pacific Grove, CA, USA~\cite{schmid_optim_2024}.}%
\thanks{L. Schmid and L. Schmalen are with the Communications Engineering Lab (CEL), Karlsruhe Institute of Technology (KIT), Hertzstr. 16, 76187 Karlsruhe, Germany (e-mail: \texttt{first.last@kit.edu}). T. Raviv and N. Shlezinger are with the School of ECE, Ben-Gurion University of the Negev, Be'er-Sheva, Israel (e-mail: \texttt{tomerraviv95@gmail.com}, \texttt{nirshl@bgu.ac.il}).}%
}

\markboth{Submitted version, \today}%
{Submitted version, \today}

\maketitle

\begin{abstract}
We investigate the application of the factor graph framework for blind joint channel estimation and symbol detection on time-variant linear inter-symbol interference channels.
In particular, we consider the expectation maximization (EM) algorithm for maximum likelihood estimation, which typically suffers from high complexity as it requires the computation of the symbol-wise posterior distributions in every iteration.
We address this issue by efficiently approximating the posteriors using the belief propagation (BP) algorithm on a suitable factor graph. By interweaving the iterations of BP and EM, the detection complexity can be further reduced to a single BP iteration per EM step.
In addition, we propose a data-driven version of our algorithm that introduces momentum in the BP updates and learns a suitable EM parameter update schedule, thereby significantly improving the performance-complexity tradeoff with a few offline training samples.
Our numerical experiments demonstrate the excellent performance of the proposed blind detector and show that it even outperforms coherent BP detection in high signal-to-noise scenarios. 
\end{abstract}

\acresetall

\IEEEpeerreviewmaketitle

\section{Introduction}
\IEEEPARstart{W}{e} study the fundamental problem of symbol detection in digital communications, and particularly the inference of transmitted symbols at the receiver impaired by a linear channel with memory and \ac{AWGN}.
While there exists a plethora of detection algorithms ranging from low-complexity linear detectors~\cite{proakis_digital_2007} to high-performance solutions that are optimal concerning the symbol error probability~\cite{bahl_optimal_1974}, the majority of works in the literature assume the availability of \ac{CSI} at the receiver. In most practical systems, this knowledge of the underlying channel is gained by sending predesigned \emph{training} or \emph{pilot} symbols from which the receiver can estimate the current channel state. The transmission of such training symbols reduces the data rate which leads, especially for rapidly time-variant channels, to a substantial reduction of the communication throughput~\cite{tong_multichannel_1998}.
For example, in many emerging communication scenarios with short block length transmission, like low-latency communications in 5G/6G or Internet of Things systems, the training signals occupy a significant part of the short transmission blocks.
Hence, there is considerable interest in efficient blind detection schemes that do not rely on the availability of \ac{CSI}.

State-of-the-art blind detectors like the \ac{CMA}~\cite{godard_self-recovering_1980} or variational methods~\cite{caciularu_unsupervised_2020, lauinger_blind_2022} are still consistently outperformed by coherent detection schemes, i.e., detectors with fully available \ac{CSI}.
On the other hand, there exist blind estimation and detection algorithms based on the \ac{ML} criterion~\cite{ghosh_maximum-likelihood_1992, kaleh_joint_1994, tong_multichannel_1998} that are competitive with coherent detectors in terms of detection performance. For instance, a joint estimation and detection approach based on the Baum-Welch algorithm is proposed in \cite{kaleh_joint_1994}, which is an earlier version of the \ac{EM} algorithm. The \ac{EM} algorithm aims to iteratively converge to a local \ac{ML} solution by alternately performing likelihood-increasing detection and estimation steps that mutually depend on each other~\cite{bishop_pattern_2006}.
However, such iterative estimation and detection schemes typically induce a notable computational burden, e.g., the method proposed in \cite{kaleh_joint_1994} requires applying a forward-backward algorithm for the \ac{MAP} detection in every iteration. 

An alternative approach to design receivers that can operate without \ac{CSI} is based on machine learning. %
The ability of \acp{DNN} to efficiently approximate high-dimensional non-linear mappings has led to a new class of so-called ``deep receivers'' that do not (only) rely on the underlying channel model but operate in a data-driven fashion~\cite{dai2020deep}. For instance, the authors in \cite{dorner_learning_2023} use a Turbo-autoencoder neural network to learn a holistic transmission scheme with superimposed pilots for joint detection and decoding, enabling efficient short-packet communications.  On the other hand, the approach in \cite{farsad_neural_2018} is completely channel-agnostic, i.e., the communication channel is assumed to be completely unknown and a recurrent neural network is trained to perform the detection task. In comparison to ``classical'' receivers that use training symbols to estimate certain parameters of the underlying channel model, a deep receiver instead uses the available data to directly adapt the \ac{DNN}-based detector to the channel.

To reduce the required number of training symbols, model-based deep learning \cite{shlezinger_model-based_2023, he_model-driven_2019} combines the advantages of both methods---model-based domain knowledge as well as data-driven optimization---by either integrating knowledge of the underlying model in the \ac{DNN} design, or vice versa, by replacing certain building blocks of an otherwise model-based algorithm with \acp{DNN}~\cite{shlezinger2023model}. Thereby, an algorithm can be significantly improved without sacrificing the benefits of being model-based.
In \cite{shlezinger_viterbinet_2020}, \acp{DNN} are integrated into the well-known Viterbi algorithm to replace all \ac{CSI}-dependent computations. %
Another example is BCJRNet~\cite{shlezinger_data-driven_2020}, which builds upon the factor graph model of the BCJR algorithm~\cite{bahl_optimal_1974} and replaces the factor nodes in the graph with a mapping that is learned from a small training set. 
While these data-aided receivers do not require \ac{CSI}, they still rely on labeled data obtained from, e.g., pilots. Moreover, their reliance on \acp{DNN} makes their online adaptation to rapidly time-variant channels challenging in terms of efficient learning~\cite{raviv2023online} and data accumulation~\cite{raviv2023data}, see also~\cite{he_model-driven_2019,raviv_adaptive_2023}.

In this work we propose a blind receiver that is particularly designed to efficiently operate in rapidly time-variant linear \ac{ISI} channels without requiring pilots. We design our receiver in two stages.
We first propose a novel joint channel estimation and symbol detection scheme based on a combination of the \ac{EM} algorithm and \ac{BP} on factor graphs that alleviates the necessity of pilots of coherent detectors and the extensive computational complexity of other non-coherent methods. 
Based on the so-called Ungerboeck observation model \cite{ungerboeck_adaptive_1974}, we use a dedicated factor graph on which the \ac{BP} algorithm efficiently implements symbol detection with particularly low complexity, namely, with linear complexity in the number of symbols \emph{and} in the memory of the channel~\cite{colavolpe_siso_2011}.
Dealing with continuous variables like channel parameters in factor graphs is typically challenging because the application of \ac{BP} usually leads to intractable expressions that must be approximated, e.g., using Gaussian approximation or quantization methods~\cite{liu_joint_2009}. To overcome this issue, Eckford proposed the integration of the \ac{EM} algorithm into the factor graph framework and showed its appealing property of breaking cycles in the graph~\cite{eckford_channel_2004}.
The idea was later formalized as a local message passing algorithm in~\cite{dauwels_expectation_2005} and applied to practical communication systems such as~\cite{wang_joint_2014}.
Inspired by this idea, we integrate our \ac{BP}-based detector into the \ac{EM} algorithm to form a joint estimation and detection framework. In contrast to~\cite{dauwels_expectation_2005} and~\cite{wang_joint_2014}, we do not use the \ac{EM} algorithm as a local message update rule, but instead perform global \ac{EM} parameter updates, into which the \ac{BP} algorithm is integrated. This turns out to be more efficient in terms of complexity, since we can derive closed-form update equations for each individual EM parameter, respectively.
Moreover, to avoid running the full detection algorithm after each \ac{EM} step, we leverage the iterative nature of factor graph-based message passing and interweave the \ac{EM} estimation steps with the \ac{BP} message passing iterations. 
Thereby, the total number of required \ac{BP} iterations for blind joint estimation and detection does not substantially increase compared to coherent detection. This leads to a significant complexity reduction compared to other approaches in literature, e.g., \cite{wang_joint_2014}, where each \ac{EM} iteration requires multiple \ac{BP} iterations.

A known weakness of the \ac{EM} algorithm is its sensitivity to initialization~\cite{bishop_pattern_2006,shireman_examining_2017}. While most works in the literature circumvent this problem by using pilot symbols to obtain an initial channel estimate, we address this challenge for the blind estimation scenario and propose a novel initialization method that leverages a lightweight version of the \ac{VAE}.

Next, we follow the concept of model-based machine learning and extend our blind factor graph-based estimation and detection
scheme into a data-driven receiver. In line with other works in literature, like~\cite{shlezinger_data-driven_2020}, we call the optimized version of the factor graph-based EMBP algorithm a \emph{data-driven factor graph}. However note that, 
compared to the approaches in~\cite{shlezinger_viterbinet_2020} and \cite{shlezinger_data-driven_2020}, 
our method is already adaptive with respect to the channel without the requirement of training symbols, thus alleviating the possibly computationally complex frequent retraining that channel agnostic deep receivers typically require to be adaptive~\cite{he_model-driven_2019}. 
Therefore, we can avoid online training and only need to perform one generic offline training prior to transmission in order to optimize the detection performance as well as the computational efficiency of our proposed algorithm for a broad range of channel states.
To this end, we unfold the iterative \ac{BP} algorithm and replace the \ac{BP} message updates with a convex combination of old and new messages. While introducing only one trainable scalar parameter per iteration, this modification significantly improves the convergence behavior and thus the detection performance. Moreover, we propose a simple, yet effective method to learn a suitable update schedule for the \ac{EM} algorithm and show that the optimized schedule can substantially reduce the number of necessary \ac{EM} steps by allowing parallel parameter updates while preserving the stability of the \ac{EM} algorithm. 

Our main contributions are summarized as follows:
\begin{itemize}
    \item We derive a novel algorithm for joint channel estimation and detection on linear \ac{ISI} channels that offers a high precision at low computational complexity based on the integration of \ac{BP} into the \ac{EM} algorithm.
    \item We significantly improve the performance-complexity tradeoff by extending our algorithm into a trainable machine learning model, that leverages offline data to 
    learn an effective update schedule for the \ac{EM} algorithm and to improve the \ac{BP}-based detection performance via learned weighting factors.
    \item We extensively evaluate the estimation and detection performance of the proposed EMBP scheme in numerical simulations with various transmission scenarios. In particular, we analyze the sensitivity of the \ac{EM} algorithm with respect to different initialization methods and compare the detection performance with other coherent and blind detection schemes from literature, including both model-based and data-driven algorithms.  
    As a remarkable feature, we show that the EMBP algorithm outperforms a comparable coherent \ac{BP}-based detector by inherently estimating a ``surrogate'' channel which is better suited for the suboptimal \ac{BP} algorithm.
    Our simulations also show that the low-complexity blind EMBP scheme performs comparably to non-blind pilot-based \ac{MAP} detectors. Furthermore, we show that the EMBP algorithm can even outperform the pilot-based detectors when compared in the context of a \ac{FEC} scheme, by leveraging the saved pilot overhead for an enhanced error correction capability. 
\end{itemize}

The remainder of this paper is organized as follows: In Section~\ref{sec:channel_model_and_prelims} we present the system model, briefly review factor graph-based symbol detection, and formulate the challenges and evaluation metrics of blind symbol detection. Section~\ref{sec:EMBP} proposes the EMBP algorithm as a blind detection algorithm as well as its data-driven version EMBP$^\star$. Section~\ref{sec:experiments} presents a comprehensive numerical evaluation of the proposed algorithms and compares the estimation and detection performance to other established methods. Section~\ref{sec:conclusion} concludes the paper.

\subsection*{Notation}
Throughout the paper, we use upper case bold letters to denote matrices $\bm{X}$ with entries $X_{m,n}$ at row $m$ and column $n$. Lower case bold letters are used for column vectors $\bm{x}$. The $i$th element of $\bm{x}$ is written as $x_i$. $\lVert \cdot \rVert$ denotes the Euclidean norm, $(\cdot)^{\textrm{T}}$ is the transpose of a matrix or vector and $(\cdot)^{\textrm{H}}$ is the conjugate transpose (Hermitian) operator. For a complex number ${c\in \mathbb{C}}$, ${\text{Re}\{ c \}}$ ${(\text{Im}\{ c \})}$ denotes its real (imaginary) part and ${c^\star}$ is its complex conjugate.
The all-zeros and all-ones vectors are written as $\bm{0}$ and $\bm{1}$, respectively, and $\bm{e}_i$ denotes a vector that has a $1$ at position $i$ and zeros everywhere else.
The ceiling and flooring operation of a floating point number~$x$ are indicated by $\lceil x \rceil$ and $\lfloor x \rfloor$, respectively. ${(i~\text{mod}~j)}$ denotes the modulo operation, i.e., the remainder of Euclidean division of $i$ by $j$ with ${i,j\in\mathbb{N}}$.
The probability density function of a continuous random variable~${y}$ is denoted by $p_{{y}}(y)$ or $p(y)$ and the probability mass function of a discrete random variable $x$ is $P_{{x}}(x)$ or $P(x)$. To keep the notation simple, we do not use a special notation for random variables since it is always clear from the context.
The Gaussian distribution, characterized by its mean $\mu$ and variance $\sigma^2$, 
is written as $\mathcal{N}(\mu,\sigma^2)$.
The expected value of a random variable ${x}$ is denoted by 
$\mathbb{E}_{{x}} \mleft \{ {x} \mright \}$ and the mutual information between ${x}$ and ${y}$ is $I({x};{y})$.
We use calligraphic letters to denote a set $\mathcal{X}$ of cardinality $|\mathcal{X}|$.

\section{System Model and Preliminaries}\label{sec:channel_model_and_prelims}
\subsection{System Model}\label{subsec:system_model}
We consider the transmission over a time-variant channel in the digital baseband which is impaired by linear \ac{ISI} and \ac{AWGN}~\cite{proakis_digital_2007}. The dynamics of the communication channel are assumed to be block-fading, i.e., the channel coefficients remain constant for a block of $N$ consecutively transmitted symbols and change to an independent realization in the next block~\cite{yang_block-fading_2013}. Each transmission block contains an information sequence ${\bm{c}\in\mathcal{M}^N}$, where each symbol $c_n$ is sampled independently and uniformly from a constellation ${\mathcal{M} = \{\text{c}_i \in \mathbb{C}, i=1,\ldots,M \}}$.
The corresponding bit pattern of length ${m :=\log_2 \mleft( M \mright)}$ is denoted by ${\bm{b}(c_n)} = \left( b_1(c_n),\ldots,b_m(c_n) \right)$.
The receiver observes the sequence
\begin{equation*}
    \bm{y} = 
    \underbrace{
    \begin{pmatrix} 
    h_0   &     &  &  \\
    \vdots& h_0  & \scalebox{1.3}{$\bm{0}$} &  \\
    h_L   &\vdots& \ddots & \\
         & h_L  &        & h_0\\
     &   \scalebox{1.3}{$\bm{0}$}  & \ddots  & \vdots \\
         &      &  & h_L\\
    \end{pmatrix}
    }_{=: \bm{H}}
    \underbrace{
    \begin{pmatrix} c_1 \\ c_2 \\ \vdots \\ c_N \end{pmatrix} }_{=: \bm{c}}
    + 
    \underbrace{
    \begin{pmatrix} w_1 \\ w_2  \\ \vdots \\ \\ w_{N+L} \end{pmatrix} }_{=: \bm{w}} ,
    \label{eqn:Channel}
\end{equation*}
where ${\bm{h} = (h_0,\ldots,h_L)^{\mathrm{T}} \in \mathbb{C}^{L+1}}$ describes the impulse response of the linear \ac{ISI} channel of length ${L+1}$ and ${w_n \sim \mathcal{CN}(0,\sigma^2)}$ are independent noise samples from a complex circular Gaussian distribution. For each transmission block, ${\bm{h}:=\alpha \cdot \tilde{\bm{h}}}$ is generated by ${\tilde{\bm{h}}\sim\mathcal{CN}(\bm{0},\bm{I}_{L+1})}$ and ${\alpha=1 / \lVert \tilde{\bm{h}} \rVert}$.
The channel is thus fully characterized by the parameter vector~$\bm{\theta} := (h_0, \ldots, h_L, \sigma^2 )^{\rm T}$ of length ${L+2}$. %
We define the \ac{SNR} at the receiver as
\begin{equation*}
    \mathsf{snr} := \frac{\lVert \bm{h} \rVert^2 \cdot \mathbb{E}_{\bm{h}
    } \mleft\{ \lVert \bm{c}\rVert^2 \mright\}}{\mathbb{E}_{\bm{w}} \mleft\{ \lVert \bm{w} \rVert^2 \mright\}} = \frac{\sum\limits_{c \in \mathcal{M}}|c|^2}{M \sigma^2}.
\end{equation*}

\subsection{Factor Graph-based Symbol Detection}\label{sec:ufg}
\subsubsection{Symbol Detection}
For each block, our goal is to infer the information sequence~$\bm{c}$ from the  observation~$\bm{y}$ at the receiver. Since the symbols and the channel parameters are independently sampled, this inference task can be approached separately for each block. 
In the context of Bayesian inference, we are interested in the symbol-wise \acp{APP}
\begin{equation}\label{eq:marginalization}
    P(c_n = \text{c} |\bm{y}) = \sum_{\mathclap{\bm{c} \in \mathcal{M}^N,\, c_n = \text{c}}} P(\bm{c|y}), \quad n=1,\ldots,N.
\end{equation} 

\subsubsection{Factor Graph Representation}
Message passing on factor graphs~\cite{kschischang_factor_2001} is a powerful tool to derive efficient algorithms for marginal inference such as the problem in~\eqref{eq:marginalization}.
In order to apply the factor graph framework to the considered problem, we first need to find a suitable factorization of the joint \ac{APP} distribution $P(\bm{c}|\bm{y})$.
Applying Bayes' theorem, we obtain
\begin{equation}\label{eq:Bayes}
    P(\bm{c}|\bm{y}) = \frac{p(\bm{y}|\bm{c})P(\bm{c})}{p(\bm{y})} = \frac{p(\bm{y}|\bm{c})}{M^N p(\bm{y})} \propto p(\bm{y}|\bm{c}).
\end{equation}
The proportionality $\propto$ in~\eqref{eq:Bayes} denotes that two terms are only differing in a factor independent of $\bm{c}$.
The likelihood
\begin{equation*}
    p(\bm{y}|\bm{c}) = \frac{1}{(\pi \sigma^2)^{N}} \exp \left ( -\frac{\Vert \bm{y-Hc} \Vert ^2}{\sigma^2} \right )
\end{equation*}
can be written as
\begin{equation}
    p(\bm{y|c}) \propto \exp\mleft( {2{\text{Re}}\mleft\{ \bm{c}^{\text H} \bm{H}^{\text H} \bm{y} \mright\} -\bm{c}^{\text H}\bm{H}^{\text H}\bm{H}\bm{c}\over \sigma^2}  \mright). \label{eq:expanded_likelihood}
\end{equation}
Based on the observation model by Ungerboeck~\cite{ungerboeck_adaptive_1974}, we substitute
\begin{equation*}
    \bm{x} := \bm{H}^{\text H} \bm{y}, \qquad
    \bm{G} := \bm{H}^{\text H} \bm{H} \label{eq:x_matched_filter},
\end{equation*}
which can be interpreted as the matched filtered versions of the observation and the channel matrix, respectively. By using
\begin{align*}
    \bm{c}^{\text H}\bm{x} &= \sum\limits_{n=1}^{N} x_n c_n^{\star}, \\
    \bm{c}^{\text H}\bm{G}\bm{c} &= \sum\limits_{n=1}^{N} G_{n,n} |c_{n}|^2 - \sum\limits_{n=1}^{N} \sum\limits_{\substack{m=1 \\ m \neq n}}^{N} \text{Re} \mleft\{ G_{n,m} c_{m} c_{n} \mright\},
\end{align*}
in~\eqref{eq:expanded_likelihood}, we obtain a factorized version of the likelihood
\begin{equation}
    p(\bm{y}|\bm{c}) \propto \prod\limits_{n=1}^{N} \left [ \text{e}^{F_n(c_n) }
                \prod\limits_{m=1}^{n-1}  \text{e}^{I_{n,m}(c_n,c_m) } \right ],
\label{eq:Ungerboeck_factorization}
\end{equation}
where we introduced the local functions
\begin{align}
    F_n(c_n) &:=  \frac{1}{\sigma^2} {\text{Re}} \mleft\{ 2 x_n c_n^\star - G_{n,n}|c_n|^2 \mright\} \label{eq:f_fn} \\
    I_{n,m}(c_n,c_m) &:= -\frac{2}{\sigma^2} {\text{Re}} \mleft\{  G_{n,m} c_m c_n^\star  \mright\} . \label{eq:I_fn}
\end{align}

A factor graph visualizes a factorization, such as~\eqref{eq:Ungerboeck_factorization}, using a bipartite graph~\cite{kschischang_factor_2001}. Every variable $c_n$ is represented by a circular node, a so-called variable node. Factor nodes represent the local functions of the factorization and are visualized by square nodes. A factor node is connected with a variable node via an undirected edge if and only if the corresponding factor is a function of this variable. For the factorization in~\eqref{eq:Ungerboeck_factorization}, this results in the factor graph illustrated in Fig.~\ref{fig:factor_graph_ungerboeck}.
\begin{figure}[tb] %
\centering
\tikzstyle{fn} = [draw, very thick, regular polygon, regular polygon sides=4, minimum width = 2.5em, inner sep=0pt, rounded corners]
\tikzstyle{vn} = [draw, very thick, circle, inner sep=0pt, minimum size = 2em]
\begin{tikzpicture}[auto, node distance=3em and 3.5em, thick]
\clip (1.7, -2.1) rectangle (9.5, 0.37);
    \node [vn, draw=none] (c0){};
    \node [vn, draw=none, right= of c0] (c1){};
    \node [vn, label=center:$c_{3}$, right= of c1] (c2){};
    \node [vn, label=center:$c_{4}$, right= of c2] (c3){};
    \node [vn, label=center:$c_{5}$, right= of c3] (c4){};
    \node [vn, draw=none, right= of c4] (c5){};
    \node [fn, draw=none, right=0.5em of c4] (rdots) {$\cdots$};
    \node [vn, draw=none, right= of c5] (c6){};
    \node [fill=KITcyan!35!white, fn, label=center: $F_{3}$, left=0.5em of c2] (p2){};
    \node [fn, draw=none, left=0.5em of p2] (rdots) {$\cdots$};
    \node [fill=KITcyan!35!white, fn, label=center: $F_{4}$, left=0.5em of c3] (p3){};
    \node [fill=KITcyan!35!white, fn, label=center: $F_{5}$, left=0.5em of c4] (p4){};
    \node [fill=KITorange!50!white, fn, label=center: $I_{3,1}$, below= of c2.260, anchor = north east] (I20){};
    \node [fill=KITorange!50!white, fn, label=center: $I_{3,2}$, below= of c2.280, anchor = north west] (I21){};
    \node [fill=KITorange!50!white, fn, label=center: $I_{4,2}$, below= of c3.260, anchor = north east] (I31){};
    \node [fill=KITorange!50!white, fn, label=center: $I_{4,3}$, below= of c3.280, anchor = north west] (I32){};
    \node [fill=KITorange!50!white, fn, label=center: $I_{5,3}$, below= of c4.260, anchor = north east] (I42){};
    \node [fill=KITorange!50!white, fn, label=center: $I_{5,4}$, below= of c4.280, anchor = north west] (I43){};
    \node [fn, draw=none, below= of c5.260, anchor = north east] (I53){};
    \node [fn, draw=none, below= of c5.280, anchor = north west] (I54){};
    \node [fn, draw=none, below= of c6.260, anchor = north east] (I64){};
    \node [fn, draw=none, below= of c6.280, anchor = north west] (I65){};
    \draw[-] (c2.west) -- (p2.east);
    \draw[-] (c3.west) -- (p3.east);
    \draw[-] (c4.west) -- (p4.east);
    \draw[-] (I20.north) -- (c2.south);
    \draw[dotted] (I20.north) -- ($(I20.north)!0.4!(c0.south)$);
    \draw[-] (I20.north) -- ($(I20.north)!0.2!(c0.south)$);
    \draw[-] (I21.north) -- (c2.south);
    \draw[-] (I21.north) -- ($(I21.north)!0.58!(c1.south)$);
    \draw[dotted] (I21.north) -- ($(I21.north)!1.0!(c1.south)$);
   \draw[-] (c2.south) -- (I32.north);
    \draw[-] (c2.south) -- (I42.north);
    \draw[-] (I31.north) -- (c3.south);
    \draw[-] (I31.north) -- ($(I31.north)!0.7!(c1.south)$);
    \draw[dotted] (I31.north) -- ($(I31.north)!1.0!(c1.south)$);
    \draw[-] (I32.north) -- (c3.south);
    \draw[-] (c3.south) -- (I43.north);
    \draw[-] (c3.south) -- ($(c3.south)!0.8!(I53.north)$);
    \draw[dotted] (c3.south) -- ($(c3.south)!1.0!(I53.north)$);
    \draw[-] (I42.north) -- (c4.south);
    \draw[-] (I43.north) -- (c4.south);
    \draw[-] (c4.south) -- ($(c4.south)!0.4!(I54.north)$);
    \draw[dotted] (c4.south) -- ($(c4.south)!0.8!(I54.north)$);
    \draw[-] (c4.south) -- ($(c4.south)!0.28!(I64.north)$);
    \draw[dotted] (c4.south) -- ($(c4.south)!0.4!(I64.north)$);
\end{tikzpicture}
\caption{Factor graph representation of \eqref{eq:Ungerboeck_factorization} for a channel with memory $L=2$.}
    \label{fig:factor_graph_ungerboeck}
\end{figure}
Due to the band structure of $\bm{G}$, many of the factor nodes $I_{n,m}(c_n,c_m)$ represent the identity function and can thus be omitted, leading to a sparse graph representation.
Note that there exist different factorizations of $p(\bm{y}|\bm{c})$, which consequently lead to distinct graphs~\cite{schmid_neural_2022}. 
However, the specific structure of the factor graph in Fig.~\ref{fig:factor_graph_ungerboeck}, originally proposed by Colavolpe et al.\ in~\cite{colavolpe_siso_2011}, is particularly favorable in terms of computational complexity in the context of marginal inference based on the \ac{BP} algorithm, which we  discuss in the following.

\subsubsection{Belief Propagation}
The sum-product algorithm, also known as \ac{BP}, is a message passing algorithm that operates in a graph and attempts to determine the marginalization towards each variable in the factor graph, respectively~\cite{kschischang_factor_2001}. 
To this end, messages are propagated within the factor graph along its edges.
In contrast to tree-structured graphs, where the messages can travel through the entire graph in a single forward and backward path, message passing in graphs with cycles yields an iterative algorithm, i.e., the messages keep traveling within the cyclic graph until some stopping criterion is fulfilled, e.g., 
a maximum number of message passing iterations is reached.

For the graph in Fig.~\ref{fig:factor_graph_ungerboeck}, we denote the message\footnote{For reasons of numerical stability and practicability, we describe the messages and their updates in the logarithmic domain.} sent in iteration~${t\in \mathbb{N}}$ from a variable node $c_n$ to a factor node $I_{n,m}(c_n,c_m)$ with ${n>m}$ as $\mu^{(t)}_{n,m}(c_n)$\footnote{Note the distinction to the message $\mu^{(t)}_{n,m}(c_m)$ which originates from variable node~$c_m$, indicated by the function argument.}. The message traveling on the same edge but in the opposite direction is denoted as $\nu^{(t)}_{n,m}(c_n)$.
All messages are initialized with an unbiased state, e.g., ${\mu^{(0)}_{n,m}(c_n) = -\log M}$. 
Then, $T$ message passing iterations are performed. Each iteration~$t$ consists of the parallel update of all messages from variable to factor nodes
\begin{subequations} \label{eq:VN_update}
\begin{align}
   \label{eq:VN_updateA}
    \mu_{n,m}^{(t)}(c_n) &= F_n(c_n) + \sum\limits_{\mathclap{\substack{m'=1 \\ m' \neq m}}}^{n-1} \nu_{n,m'}^{(t-1)}(c_n) + \sum\limits_{\mathclap{n'=n+1}}^N \nu_{n',n}^{(t-1)}(c_n), \\
    \mu_{n,m}^{(t)}(c_m) &= F_m(c_m) + \sum\limits_{\mathclap{m'=1}}^{m-1} \nu_{m,m'}^{(t-1)}(c_m) + \sum\limits_{\mathclap{\substack{n'=m+1 \\ n' \neq n}}}^N \nu_{n',m}^{(t-1)}(c_m),
\end{align}
\end{subequations}
followed by the parallel update of all messages from factor to variable nodes
\begin{subequations} \label{eq:FN_update}
\begin{align}
    \nu_{n,m}^{(t)}(c_n) &= \log \sum_{c_m} \exp \mleft( I_{n,m}(c_n,c_m) + \mu^{(t)}_{n,m}(c_m) \mright), \\
    \nu_{n,m}^{(t)}(c_m) &= \log \sum_{c_n} \exp \mleft( I_{n,m}(c_n,c_m) + \mu^{(t)}_{n,m}(c_n) \mright). 
\end{align}
\end{subequations}
The final \ac{BP} result is obtained after the last message passing iteration ${t=T}$ by combining all incident messages at a variable node to the so-called beliefs via
\begin{equation}
    b_n(c_n) = \gamma_n \exp \mleft( F_n(c_n) + \sum\limits_{\mathclap{m'=1}}^{n-1} \nu_{n,m'}^{(T)}(c_n) + \sum\limits_{\mathclap{n'=n+1}}^N \nu_{n',n}^{(T)}(c_n) \mright), \label{eq:beliefs}
\end{equation}
where the $\gamma_n \in \mathbb{R}$ are chosen such that $b_n(c_n)$ are normalized probability distributions.
The beliefs are an approximation of the symbol-wise \ac{APP} distributions $P(c_n |\bm{y}) \approx b_n(c_n)$ and the quality of this approximation strongly depends on the number of iterations~$T$ as well as on the nature of the underlying factor graph. For the specific graph in Fig.~\ref{fig:factor_graph_ungerboeck}, the approximation quality thus varies with the block length $N$, the channel parameters $\bm{\theta}$ and with the observation~$\bm{y}$.

A major benefit of \ac{BP} on a suitable factor graph is its ability to approximate the \ac{APP} with low computational complexity. The asymptotic complexity of the described algorithm is ~${\mathcal{O}(TNLM^2)}$~\cite{colavolpe_siso_2011}. In comparison, exact inference implemented by the BCJR algorithm~\cite{bahl_optimal_1974} scales with~${\mathcal{O}(NM^{L+1})}$, i.e., the complexity grows exponentially with the memory of the channel ${L}$ instead of linearly.

\subsection{Problem Formulation} \label{sec:problem_formulation}
We consider symbol detection without \ac{CSI}, i.e., the parameters $\bm{\theta}$ are unknown to both the transmitter and the receiver. 
An unsupervised estimate of the parameters can be found by maximizing the likelihood of the observation $\bm{y}$ with respect to the channel parameters
\begin{equation}
    \hat{\bm{\theta}} = \arg\max_{\bm{\theta}} p(\bm{y}|\bm{\theta}),
    \label{eq:exact_ML}
\end{equation}
where we have included the unknown parameters $\bm{\theta}$ as latent variables in the model. Once $  \hat{\bm{\theta}}$ is estimated, it can be used for \ac{CSI}-based symbol detection. 
However, the exact \ac{ML} solution in~\eqref{eq:exact_ML} is infeasible for reasonable block lengths ${N \gg 1}$.
Nonetheless, leveraging intermediate results from symbol detection for the estimation task can help to reduce the complexity significantly. 
This motivates our goal to design an efficient algorithm for \emph{joint} channel estimation and symbol detection. 
In this context, we are facing two major challenges:
\begin{enumerate}[label={C\arabic*}]
    \item \label{itm:performance} To achieve a good detection performance that is comparable to a coherent detection with available \ac{CSI}.
    \item \label{itm:complexity} To keep the computational complexity low such that it can be implemented on a receiver with sufficiently low power consumption and/or delay.
\end{enumerate}

\subsection{Evaluation Metrics} \label{subsec:objective_functions}
Suitable objective functions for \emph{joint} estimation and detection algorithms can either evaluate the detection or the estimation performance.
The quality of an estimated channel impulse response $\hat{\bm{h}}$ can be quantified by the squared error
\begin{equation*}
    \lVert \hat{\bm{h}} - \bm{h} \rVert^2 = \sum_{\ell=0}^L \big| \hat{h}_\ell-h_\ell \big|^2
\end{equation*}
to the ground truth~$\bm{h}$. %
For statistics over multiple channels and their respective estimations, we will consider the median and the \ac{MSE}.

As a figure of merit for the detection performance, we consider the maximum achievable information rate between the channel input and the detector output. %
Many practical transmission systems use \ac{BICM}, which decouples the 
symbol detection from a binary soft-decision \ac{FEC}~\cite{Fabregas_foundations_2008}.
In a \ac{BICM} system, the symbol detector soft output $P(c_n|\bm{y},\bm{\theta})$ is converted by a \ac{BMD} to binary soft information 
\begin{equation*}
    P(b_i(c_n) = b | \bm{y},\bm{\theta}) = \sum\limits_{\text{c} \in \mathcal{M}_i^{(b)}} P(c_n = \text{c} | \bm{y},\bm{\theta}), \quad b \in \{ 0,1 \}
\end{equation*}
with
$\mathcal{M}_i^{(\text{b})} := \mleft \{ \text{c} \in \mathcal{M} : b_{i}(\text{c}) = \text{b} \mright \}$.
The resulting bit-wise \acp{APP} are typically expressed in \acp{LLR}
\begin{equation*}
    L_{n,i}(\bm{y}) := \log \mleft( \frac{P(b_i(c_n) = 0 | \bm{y},\bm{\theta})}{P(b_i(c_n) = 1 | \bm{y},\bm{\theta})} \mright).
\end{equation*}
After interleaving, the \acp{LLR} are fed to a bit-wise soft-decision \ac{FEC}.
By interpreting the \ac{BMD} as a mismatched detector, the \ac{BMI} is an achievable information rate for \ac{BICM}~\cite{Fabregas_foundations_2008}.
The calculation of the \ac{BMI}, detailed in~\cite{alvarado_achievable_2018}, 
considers the \ac{BMD} by assuming $m$ parallel sub-channels 
transmitting on a binary basis
instead of one symbol-based channel. Assuming independent transmit bits,
the \ac{BMI} is defined as the sum of mutual informations\footnote{The mutual information is a measure between two random variables. We avoid a distinct notation for random variables as it is clear from the context.}
$I(b_i ; y)$ of $m$ unconditional bit-wise channel transmissions:
\begin{equation*}
    \text{BMI} := \sum _{i=1}^{m}\! I(b_i(c) ; y).
\end{equation*}
By a sample mean estimate over $D$ labeled data tuples, a feasible approximation is given by~\cite{alvarado_achievable_2018}
\begin{align*}
    &\text{BMI} \approx \log_2 \mleft( M \mright) - \\
    &\frac{1}{D N} \sum\limits_{\mathclap{i=1}}^{m} 
    \sum\limits_{n=1}^{N}
    \sum\limits_{d=1}^D
    \log_2 \mleft (  \exp\mleft(-(-1)^{b_{n,i}(c_n^{({d})})} L_{n,i}(\bm{y}^{(d)})\mright) + 1  \mright ), \nonumber
\end{align*} %
where $c_n^{({d})}$ and $\bm{y}^{(d)}$ are from a labeled data set~$\mathcal{D}$, which will be formally defined later in~\eqref{eqn:DataSet} in Sec.~\ref{sec:optimization}.

\section{Factor Graph-based Joint Estimation and Detection using EM and BP} \label{sec:EMBP} %

We propose a joint estimation and detection scheme for the problem formulated in Sec.~\ref{sec:problem_formulation}.
First, we derive closed-form update expressions for the \ac{EM}-based parameter estimation in Sec.~\ref{sec:EM} and show how the \ac{BP} algorithm can be applied to effectively reduce the computational complexity.
As this formulation on its own may not successfully cope with \ref{itm:performance}-\ref{itm:complexity} in short coherence durations, we empower our method in two aspects: 
$(i)$ we incorporate dedicated initialization methods, as detailed in Sec.~\ref{sec:init}; and $(ii)$ we optimize the algorithm by introducing a small set of key parameters within the algorithm that are tuned in a data-driven fashion as a model-based machine learning model in offline training~\cite{shlezinger2022model}, as detailed in Sec.~\ref{sec:optimization}.

\subsection{Expectation Maximization} \label{sec:EM}
The \ac{EM} algorithm is an established technique for finding \ac{ML} parameter estimates in probabilistic models that contain latent variables~\cite{dempster_maximum_1977}.
The basic idea is to maximize a simple lower bound of the likelihood in~\eqref{eq:exact_ML} in a 2-stage iterative algorithm.
To this end, the log-likelihood function is decomposed into~\cite{bishop_pattern_2006}
\begin{align}
    &\log p(\bm{y}|\bm{\theta}) = \log p(\bm{c},\bm{y}|\bm{\theta}) - \log P(\bm{c}|\bm{y},\bm{\theta}) \nonumber \\
    &= \underbrace{\sum\limits_{\bm{c}} Q(\bm{c}) \log \mleft( \frac{p(\bm{c},\bm{y}|\bm{\theta})}{Q(\bm{c})} \mright)}_{=: \mathcal{L}(Q,\bm{\theta})} \underbrace{- \sum\limits_{\bm{c}} Q(\bm{c}) \log \mleft( \frac{P(\bm{c}|\bm{y},\bm{\theta})}{Q(\bm{c})} \mright)}_{=: D_\text{KL}(Q \Vert P)}, \label{eq:likelihood_decomposition}
\end{align}
where $Q(\bm{c})$ can be any normalized trial distribution of $\bm{c}$. 
Jointly optimizing the log-likelihood in~\eqref{eq:likelihood_decomposition} with respect to $\bm{\theta}$ and \emph{all} possible $Q(\bm{c})$ is typically infeasible.
Due to the non-negativity of the \ac{KL} divergence~${D_\text{KL}(Q \Vert P) \geq 0}$, the functional ${\mathcal{L}(Q,\bm{\theta})}$ is a so-called \ac{ELBO}, satisfying
\begin{equation}
    \log p(\bm{y}|\bm{\theta}) \geq \mathcal{L}(Q,\bm{\theta}). \nonumber
    \label{eq:ELBO}
\end{equation}
The \ac{EM} algorithm maximizes the \ac{ELBO} in multiple consecutive steps. Starting with an initial guess for the parameters~$\hat{\bm{\theta}}^{(0)}$, the \ac{EM} algorithm iteratively performs
\begin{enumerate}
    \item \textbf{E-step}: Fix $\hat{\bm{\theta}}^{(t-1)}$ and maximize~$\mathcal{L}(Q,\bm{\theta})$ w.r.t. $Q$:
        \begin{align}
            Q &= \arg\max\limits_{Q} \mathcal{L}(Q,\bm{\theta}) \Big|_{\bm{\theta} = \hat{\bm{\theta}}^{(t-1)}} \nonumber \\
            &= P(\bm{c}|\bm{y},\hat{\bm{\theta}}^{(t-1)}). \label{eq:E-step}
        \end{align}
    \item \textbf{M-step}: Fix $Q$ and maximize~$\mathcal{L}(Q,\bm{\theta})$ w.r.t. $\bm{\theta}$:
        \begin{align}
            \hat{\bm{\theta}}^{(t)}
            &= \arg\max\limits_{\bm{\theta}} \mathcal{L}(Q,\bm{\theta}) \Big|_{Q = P(\bm{c}|\bm{y},\hat{\bm{\theta}}^{(t-1)})} \nonumber \\
            &= \arg\max\limits_{\bm{\theta}} \mathbb{E}_{ P(\bm{c}|\bm{y},\hat{\bm{\theta}}^{(t-1)})} \left\{ \log p(\bm{c},\bm{y}|\bm{\theta}) \right\} \nonumber \\
            &=: \arg\max\limits_{\bm{\theta}} \tilde{Q}\mleft(\bm{\theta} \Big\vert \hat{\bm{\theta}}^{(t-1)} \mright). \label{eq:M-step}
        \end{align}
\end{enumerate}
Despite the fact that both the \Estep\ and the \Mstep\ only maximize a lower bound of the likelihood, respectively, the
\ac{EM} algorithm has the appealing property of monotonically increasing the log-likelihood in every iteration~\cite{eckford_channel_2004}:
\begin{equation*}
    \log p(\bm{y}|\hat{\bm{\theta}}^{(t)}) \geq \log p(\bm{y}|\hat{\bm{\theta}}^{(t-1)}).
\end{equation*}
For a more detailed derivation of~\eqref{eq:E-step} and~\eqref{eq:M-step} and an  in-depth treatment of the \ac{EM} algorithm, we refer the reader to~\cite{bishop_pattern_2006}.

For the application of the \ac{EM} algorithm to our problem in~\eqref{eq:exact_ML}, we need to solve the optimization problems of the \Estep\ and \Mstep, respectively.
The optimal solution for the \Estep\ is given in~\eqref{eq:E-step} by the \ac{APP} distribution ${Q(\bm{c})=P(\bm{c}|\bm{y},\hat{\bm{\theta}}^{(t-1)}) = \prod_{n=1}^N P(c_n|\bm{y},\hat{\bm{\theta}}^{(t-1)})}$. %
Finding the symbol-wise posteriors $P(c_n|\bm{y},\hat{\bm{\theta}}^{(t-1)})$ based on the channel observation~$\bm{y}$ and some model assumptions~$\hat{\bm{\theta}}^{(t-1)}$ is a probabilistic inference task similar to the problem discussed in Sec.~\ref{sec:ufg}. In other words, the \Estep\ of the \ac{EM} algorithm resolves to  soft-output symbol detection using the \ac{CSI}~$\hat{\bm{\theta}}^{(t-1)}$.

Driven by the requirement of finding a \emph{low-complexity} solution, we propose to use the factor graph-based \ac{BP} algorithm discussed in Sec.~\ref{sec:ufg} to efficiently perform the \Estep\ by approximating the symbol-wise posteriors $P(c_n|\bm{y},\hat{\bm{\theta}}^{(t-1)})$ with the message passing beliefs $b_n(c_n)$.

Based on the \ac{APP} distribution $P(\bm{c}|\bm{y},\hat{\bm{\theta}}^{(t-1)})$ or an approximation thereof, the \Mstep\ updates the parameter estimates~$\hat{\bm{\theta}}^{(t-1)}$. 
The optimization problem in~\eqref{eq:M-step} cannot be solved jointly for the entire parameter vector~$\bm{\theta}$ in closed form. However, we can find a closed-form solution for the maximization of each scalar parameter $\theta_\ell$ by solving ${{\left. \dfrac{\partial}{\partial \theta_\ell} \tilde{Q}\mleft(\bm{\theta} \Big\vert \hat{\bm{\theta}}^{(t-1)} \mright)\right|_{\theta_\ell = \hat{\theta}_{\ell}^{(t)}} } = 0}$, as stated in the following theorem:
\begin{theorem}
\label{thm:EMSolution}
The setting of $\hat{\bm{\theta}}^{(t)}=(\hat{h}_0, \ldots, \hat{h}_L, \hat{\sigma}^2 )^{\rm T}$ which solves the optimization problem in~\eqref{eq:M-step} for the system model detailed in Sec.~\ref{subsec:system_model} along each dimension of $\bm{\theta}$ independently is given by~\eqref{app:eq:sigma2} and \eqref{app:eq:update_hcomplex}, where $B$, $C(c_n,\bm{h})$ and $D(c_n,c_m,\bm{h})$ are given in~\eqref{app:eq:BCD}.
\end{theorem}
\begin{figure*}
\begin{align}
&\begin{aligned}
      \hat{\sigma}^{2(t)} = \frac{1}{N} \Bigg[ &B - \sum\limits_{n=1}^N \sum\limits_{c_n} P(c_n|\bm{y},\hat{\bm{\theta}}^{(t-1)}) \bigg( C(c_n,\bm{h}^{(t-1)}) - \sum\limits_{m<n} \sum\limits_{c_m} P(c_m|\bm{y},\hat{\bm{\theta}}^{(t-1)}) \, D(c_n,c_m,\bm{h}^{(t-1)}) \bigg) \Bigg], 
\end{aligned} \label{app:eq:sigma2} \\
&\begin{aligned}
     \hat{h}_{\ell}^{(t)} 
     = \Bigg( &\sum\limits_{n=1}^N \sum\limits_{c_n} P(c_n|\bm{y},\hat{\bm{\theta}}^{(t-1)})  y_{n+\ell} c_n^\star - \sum\limits_{\substack{k=0 \\ k \neq \ell}}^L \sum\limits_{c_{n-|\ell-k|}} P(c_{n-|\ell-k|}|\bm{y},\hat{\bm{\theta}}^{(t-1)}) \, h_k^{(t-1)} \\
     &\bigg( \text{Re}\{ c_{n-|\ell-k|} c_n^\star \}
     - \text{j} \cdot \text{Im}\{ c_{n-|\ell-k|} c_n^\star \} \text{sign} \mleft\{ \ell-k \mright\} \bigg) \Bigg) 
     \cdot \left( \sum\limits_{n=1}^N \sum\limits_{c_n} P(c_n|\bm{y},\hat{\bm{\theta}}^{(t-1)}) |c_n|^2 \right)^{-1} 
\end{aligned} \label{app:eq:update_hcomplex} %
\end{align}
\hrule
\end{figure*}
\begin{IEEEproof}
    The proof is given in Appendix \ref{app:sec:EM}. 
\end{IEEEproof}
\smallskip

Theorem~\ref{thm:EMSolution} implies that the update of a single parameter $\hat{\theta}_{\ell}^{(t)}$ 
does not directly depend on its respective predecessor $\hat{\theta}_{\ell}^{(t-1)}$. This dependency is merely given indirectly via the posterior distribution $P(\bm{c}|\bm{y},\hat{\bm{\theta}}^{(t-1)})$ in the \Estep.
On the other hand, the update equations directly depend on the estimates of other parameters, e.g., the update of~$\hat{h}_\ell^{(t)}$ in~\eqref{app:eq:update_hcomplex} is a function of $\hat{h}_k^{(t-1)}$ with ${k\neq \ell}$.
This suggests that for optimal parameter updates, the \Mstep\ should only update one element of $\hat{\bm{\theta}}$ at a time and perform another \Estep\ before updating the next element. In this case, Theorem~\ref{thm:EMSolution} guarantees the optimality of the respective parameter update in the \Mstep. However, the element-wise parameter updates come with the price of potentially more \ac{EM} steps and thus a higher computational complexity. 
In general, we select the parameters that are updated in parallel during one particular \Mstep~$t$ by defining a sub-vector ${\hat{\bm{\theta}}_{\subset}^{(t)} := (\hat{\theta}_{J_1}, \ldots, \hat{\theta}_{J_{K^{(t)}}})^{\rm T}}$ with ${\{J_1,\ldots, J_{K^{(t)}} \} \subseteq \{ 1,\ldots, L+2 \}}$ and only updating the $K^{(t)}$ elements ${\hat{\bm{\theta}}_{\subset}^{(t)} \leftarrow \hat{\bm{\theta}}_{\subset}^{(t-1)}}$ while leaving the remaining parameter estimates unchanged.

We summarize the proposed \ac{EM} procedure for joint estimation and detection in Alg.~\ref{alg:EMBP}. The algorithm accepts the channel observation~$\bm{y}$ as well as an initialization for the parameters~$\hat{\bm{\theta}}^{(0)}$ as inputs. 
Then, it performs $T$ \ac{EM} steps, each consisting of one \ac{BP} message passing iteration in the \Estep\ followed by the update of a selection $\hat{\bm{\theta}}_{\subset}^{(t-1)}$ of parameter estimates in the \Mstep. Note that multiple \ac{BP} iterations per \Estep\ can be performed by skipping the parameter updates in one or multiple consecutive \ac{EM} steps, i.e., by setting the update vectors~$\hat{\bm{\theta}}_{\subset}^{(t-1)}$ to length~${K^{(t)}=0}$ in the respective \ac{EM} steps~$t$. The algorithm outputs a final \ac{CSI} estimate~$\hat{\bm{\theta}}$ and the beliefs ${b_n(c_n),\, n=1,\ldots,N}$ as the result of symbol detection. Because the proposed joint estimation and detection algorithm performs \ac{BP} in a factor graph for the \Estep\ of the \ac{EM} procedure, we denote it \emph{EMBP~algorithm}. 

\begin{algorithm}[t] %
    \DontPrintSemicolon
    \KwData{$\bm{y}$, $\hat{\bm{\theta}}^{(0)}$} %
    $\nu^{(0)}_{n,m}(c_k) \gets -\log M, \; k = m,n$ \tcp*{BP init.}
    \For{$t = 1,\ldots,T$}
    {
        \textbf{E-step}: BP-based symbol detection \\
        \quad $\mu_{n,m}^{(t)}(c_k), \; k=m,n \gets$ BP msg. updates~\eqref{eq:VN_update} \\
        \quad $\nu_{n,m}^{(t)}(c_k), \; k=m,n \gets$ BP msg. updates~\eqref{eq:FN_update}\\
        \quad $b_n(c_n)^{(t)} \gets$ Compute beliefs~\eqref{eq:beliefs}\\
        \textbf{M-step}: Select $K^{(t)}$ parameters $\hat{\bm{\theta}}_{\subset}^{(t-1)}$ to update \label{alg:EMschedule} \\
        \quad $\hat{\bm{\theta}}_{\subset}^{(t)} \gets$ update $\hat{\bm{\theta}}_{\subset}^{(t-1)}$ using~\eqref{app:eq:sigma2},  \eqref{app:eq:update_hcomplex} based on \\
        \nonl \quad the approximation ${b_n^{(t)}(c_n) \approx P(c_n|\bm{y},\hat{\bm{\theta}}^{(t-1)})}$ 
    }
    \KwResult{${\hat{\bm{\theta}}=\hat{\bm{\theta}}^{(T)}}$, ${b_n(c_n) = b_n^{(T)}(c_n)}$}
    \caption{EMBP} %
    \label{alg:EMBP}
\end{algorithm}

\subsection{Complexity}
\begin{table*}[t]
    \centering
\caption{Number of additions (ADD), multiplications (MULT), and $\log\Sigma\exp$-computations in the steps of the EMBP algorithm compared to other algorithms}
\begin{tabular}{l  l  l  l l}
  \toprule
    Operation & ADD (real-valued) & MULT (real-valued) & $\log\Sigma\exp$ & asymptotic complexity \\ 
  \midrule			
   \multirow{2}{*}{\Estep $\big($BP$\big)$}  & $NM\big(2LM+4L+4\big)+N\big(4L+4\big)$ & $NM\big(3\big)+N\big(4L+4\big)$  & $NM\big(2LM+1\big)$ & $\mathcal{O}\big(NLM^2\big)$ \\
                                 & $+L\big(M^2+2L+3\big)+1$ \vspace{0.7em}               & $+L\big(3M^2+2L+4\big)+M+2$         &                & \\
   \multirow{2}{*}{\Mstep $\big(\hat{\sigma}^2\big)$ } & $NM\big(2LM+4\big)+N\big(4L+6\big)$    & $NM\big(3LM+3\big)+N\big(4L+6\big)$ & - & $\mathcal{O}\big(NLM^2\big)$ \\
                                                & $+L\big(2L+5\big)+1$ \vspace{0.7em}                  & $+L\big(2L+6\big)+M+2$              &  & \\
   M-step $\big(\hat{h}_\ell\big)$                        & $NM\big(4L+6\big)$                     & $NM\big(6L+9\big)$           & - & $\mathcal{O}\big(NLM\big)$ \vspace{0.7em} \\
   VAE-LE-step & $ 12N(L+1) + 7NM + 18N $ & $ 13N(L+1) + 9NM +14N $ & - & $\mathcal{O}\big(NL + NM  \big)$ \vspace{0.7em} \\
   MAP (BCJR) & $N M^{L+1} \big( 2L+8 \big)$ & $NM^{L+1}(4L+9)$ & $M^{L+1}(3N+2L)$ & $\mathcal{O}\big( NLM^{L} \big)$ \\
  \bottomrule
\end{tabular}
\label{tab:complexity}
\end{table*}
We consider the complexity of the proposed EMBP algorithm based on Table~\ref{tab:complexity}, which provides the number of required operations for the different steps of the EMBP algorithm. For detailed remarks concerning Table~\ref{tab:complexity}, we refer the interested reader to Appendix~\ref{app:complexity}.
To obtain the overall complexity of the EMBP algorithm for a given transmission scenario, we can accumulate the executed operations of the required E/M-steps in each iteration.
If every parameter is updated a constant number of times, the asymptotic complexity of the EMBP algorithm to perform blind estimation and joint detection becomes ${\mathcal{O}\mleft( TNLM^2 + NL^2M \mright)}$, where the number of iterations~$T$ scales either constantly or linearly with $L$, depending on the concurrency of the parameter updates in the \Mstep~(line 7 in Alg.~\ref{alg:EMBP}).

\subsection{EMBP Initialization} \label{sec:init}
An important aspect of the EMBP algorithm, which we did not discuss yet, is the choice of a suitable parameter initialization~$\hat{\bm{\theta}}^{(0)}$. Initialization has been shown to substantially influence the quality of \ac{EM}-based estimation since a poor initialization might cause the \ac{EM} algorithm to converge to a locally (instead of globally) optimal solution~\cite{bishop_pattern_2006,shireman_examining_2017}.
For example, in the context of Gaussian mixture models, one of the most widely used applications of the \ac{EM} algorithm, it has been shown in~\cite{jin_local_nodate} that the EM algorithm converges with high probability to a bad local optimum which can be arbitrarily worse than that of any global optimum if the number of mixture components is 3 or larger.

For our application, the importance of a proper initialization of the EMBP algorithm
can immediately be seen if we consider the initialization of the parameter estimates $\hat{h}_\ell, \, {\ell=0,\ldots,L}$ with their expected values ${\mathbb{E}\{ h_\ell\} = 0}$. 
In this case, the EM algorithm deadlocks in the local optimum of the prior information\footnote{We assume that the constellation~$\mathcal{M}$ is axisymmetric.} ${P(\bm{c}|\bm{y},\hat{\bm{\theta}}^{(t-1)}) = P(\bm{c}) = 1/M}$ and $\hat{h}^{(t)}_\ell=0,\, \forall t$.
A more practical option is the initialization with an impulse ${\hat{\bm{h}}^{(0)} = \bm{h}_{\bm{\delta}} := \bm{e}_{\lceil L/2 \rceil}}$, which corresponds to the assumption of a memoryless channel.

We propose to initialize the estimate~$\hat{\bm{\theta}}^{(0)}$ of the EMBP algorithm by employing the \ac{VAELE}, a lightweight blind channel equalizer~\cite{lauinger_blind_2022}. 
To briefly introduce the \ac{VAELE}, we revisit the derivation of the EM algorithm in Sec.~\ref{sec:EM}. 
Tractable \ac{ML} estimation is achieved by simplifying the objective from the log-likelihood function $\log p(\bm{y}|\bm{\theta})$ to the \ac{ELBO}~$\mathcal{L}(Q,\bm{\theta})$, which is jointly maximized with respect to $\bm{\theta}$ and $Q(\bm{c})$~\cite{caciularu_unsupervised_2020}. 
The EM algorithm simplifies the problem of joint maximization by alternately fixing $\bm{\theta}$ or $Q$, which leads to the easier subproblems~\eqref{eq:E-step} and~\eqref{eq:M-step} in the \Estep\ and \Mstep, respectively.
In contrast, the idea of the \ac{VAELE} is to perform a truly \emph{joint} maximization of~$\mathcal{L}(Q,\bm{\theta})$ and instead reduce the set of possible distributions~$Q(\bm{c})$ over which the optimization is performed.
Following the idea of variational inference, the trial distribution $Q(\bm{c})$ is chosen from a family of probability distributions~$\mathcal{Q}_{\bm{\Phi}}$ which are parametrized by the so-called variational parameters~$\bm{\Phi}$~\cite{blei_variational_2017}.

In particular, the \ac{VAELE} restricts the search space $\mathcal{Q}_{\bm{\Phi}}$ to distributions $Q_{\bm{\Phi}}(\bm{c})$ which are the result of linear channel equalization using a \ac{FIR} filter with impulse response ${\bm{\Phi}\in \mathbb{C}^{L_\text{LE}+1}}$. 
To be more precise, a distribution ${Q_{\bm{\Phi}}(\bm{c})=\prod_{n=1}^N Q_{\bm{\Phi}}(c_n)}$ is obtained from the output ${\hat{\bm{c}}_\text{LE} \in \mathbb{C}^{N}}$ of the linear equalizer by applying a soft-demapping based on a Gaussian assumption:
\begin{equation*}
    Q_{\bm{\Phi}}(c_n=c) = \frac{\exp \mleft( -\frac{\lvert \hat{c}_{n,\text{LE}} - c \rvert^2}{\sigma^2_\text{VAE}} \mright)}{\sum\limits_{\mathclap{c' \in \mathcal{M}}} \exp \mleft( -\frac{\lvert \hat{c}_{n,\text{LE}} - c' \rvert^2}{\sigma^2_\text{VAE}} \mright)}, \quad c \in \mathcal{M}.
\end{equation*}

To perform blind channel equalization, the \ac{VAELE} jointly maximizes $\mathcal{L}(Q,\bm{\theta})$ with respect to the channel parameters~$\bm{\theta}$ and the taps of the linear equalization filter $\bm{\Phi}$. Typically, this maximization is based on gradient methods, e.g., applying $S_\text{VAE}$ iterations of the Adam algorithm~\cite{kingma_adam_2015}.
It is important to note that this optimization is completely unsupervised and is performed online as an integral part of the algorithm.
For more details and practical considerations we refer the reader to~\cite{lauinger_blind_2022}.
Due to its restriction to linear equalization, we presume the \ac{VAELE} to be in general less performant compared to the EMBP algorithm that employs a non-linear detector using \ac{BP} on factor graphs.
On the other hand, we expect the \ac{VAELE} to be more robust against poor initializations, since the simple structure of the \ac{FIR}-based detector is less prone to get stuck in local optima. 
In addition, the restriction to a linear filter in the equalizer leads to a particularly low complexity that scales linearly with the block length~$N$, the channel memory~$L$ and the size of the constellation~$M$ per \ac{VAE} iteration.
This is why we propose to use a few iterations~$S_\text{VAE}$ of the \ac{VAELE} to initialize the parameter estimates~$\hat{\bm{\theta}}^{(0)}$ of the EMBP algorithm.

\subsection{Factor Graph-based Deep Learning} \label{sec:optimization}

The proposed EMBP algorithm is a joint estimation and detection scheme that is strongly based on its underlying probabilistic model: the \ac{EM} algorithm aims at maximizing the likelihood~$p(\bm{y}|\bm{\theta})$ and the \ac{BP}-based symbol detection in the \Estep\ is performed in a factor graph that represents a factorization, i.e., conditional independencies among the random variables in~$p(\bm{y}|\bm{c})$. 
The reliance on such modeling brings forth several advantages, like the interpretability of intermediate results, e.g., every \ac{BP} message in the factor graph represents a marginal probability distribution. Further advantages are theoretical guarantees, for instance, the \ac{EM} algorithm provably increases the likelihood in each step.

Nonetheless, we have introduced several approximations in the derivation of the EMBP algorithm to keep it feasible in terms of complexity. First, we have approximated the true marginals $P(c_n|\bm{y},\hat{\bm{\theta}})$ with the beliefs $b_n(c_n)$ as the result of \ac{BP} on a \emph{cyclic} factor graph. Second, we have simplified the \ac{ML} parameter estimation by using the \ac{EM} algorithm as a 2-step iterative approach that maximizes the \ac{ELBO} instead of the log-likelihood. %
To overcome the performance loss that comes with these approximations and at the same time maintain a low complexity, we leverage methods of model-based deep-learning~\cite{shlezinger2022model} to develop a refined data-driven version of the EMBP algorithm in the following. 

\subsubsection{Enhancing \ac{BP} accuracy} \label{subsubsec:momentumBP}
Several approaches have been proposed in the literature to improve the performance of \ac{BP} in cyclic factor graphs, including the generalized \ac{BP} algorithm by Yedidia et al.~\cite{yedidia_generalized_2000}, and the concave-convex procedure (CCCP) which offers convergence guarantees~\cite{yuille_cccp_2002}. Importantly, these generalized message-passing algorithms can be integrated into the proposed EMBP framework, as it only requires an iterative algorithm that refines approximations of the posterior distribution. However, the aforementioned algorithms significantly increase the computational complexity which is not in line with the focus of this work.
A common method to improve the performance of \ac{BP} while maintaining its low complexity is \emph{neural BP}, which unrolls the \ac{BP} iterations in the factor graph to a feed-forward network and assigns trainable weights to every edge in the unrolled network~\cite{nachmani_deep_2018}.
Neural BP can significantly improve the detection performance for cyclic factor graphs based on the Ungerboeck observation model if the algorithm is optimized for one specific channel~\cite{schmid_low-complexity_2022}. 
However, our investigations show that it does not generalize well for a broad range of channel characteristics, as it is required for the application of blind symbol detection on block-fading channels in this work.

A much simpler generalization of the \ac{BP} algorithm, which we expect to better generalize on fading channels, is the introduction of ``momentum'' in the message updates~\cite{murphy_loopy_1999}, i.e., replacing the \ac{BP} message updates in~\eqref{eq:VN_update},~\eqref{eq:FN_update} with a convex combination of the new and the old messages, respectively. For example, the message~${\mu_{n,m}^{(t)}(c_n)}$ in iteration~$t$ is replaced with ${{\beta_\text{BP} \cdot \mu_{n,m}^{(t)}(c_n)} + {(1-\beta_\text{BP}) \cdot \mu_{n,m}^{(t-1)}(c_n)}}$.
By choosing ${0< \beta_\text{BP} < 1}$, the idea is to improve the convergence behavior of the message passing scheme compared to the original \ac{BP}~(${\beta_\text{BP}=1}$) while retaining the same fixed points of the iterative message passing.
We generalize this idea, where a constant ${\beta_\text{BP} \in (0,1]}$ is used among all message updates~\cite{murphy_loopy_1999}, by using an individual weight ${\beta_\text{BP}^{(t)} \in \mathbb{R}}$ for each \ac{BP} iteration ${t=1,\ldots,T}$, thus unfolding these iterations into a trainable machine learning architecture.
We expect that this additional degree of freedom allows the message passing to improve its convergence behavior during the early \ac{BP} iterations (global optimization) and at the same time to reduce the overall number of iterations by speeding up the local optimization during the final \ac{BP}~iterations.
Note that this momentum-based \ac{BP} is a generalization of \ac{BP} and by setting all weights ${\beta_\text{BP}=1}$, we recover the original \ac{BP} algorithm. Hence, by an optimal choice of the weights ${\beta_\text{BP}}$ with respect to a given figure of merit, the overall performance of the EMBP algorithm will only be improved.

To find the optimal weights ${\beta_\text{BP}^{(t)}}$, we use a training data set
\begin{equation}
    \label{eqn:DataSet}
    {\mathcal{D}=\{ (\bm{c}^{(d)},\bm{\theta}^{(d)},\bm{y}^{(d)}), \, d=1,\ldots,D \}},
\end{equation}
consisting of randomly and independently sampled information sequences $\bm{c}^{(d)}$, channel realizations $\bm{\theta}^{(d)}$, 
and a corresponding noisy channel observation $\bm{y}^{(d)}$.
We iteratively apply the EMBP algorithm to elements of $\mathcal{D}$, evaluate the results towards one of the objective functions introduced in Sec.~\ref{subsec:objective_functions} (\ac{MSE} or \ac{BMI}), and backpropagate its gradient with respect to the trainable parameters which are then optimized   using  Adam.

\subsubsection{Learning the \ac{EM} update schedule}
Another part of  EMBP  which we  optimize is the selection of parameters~$\hat{\bm{\theta}}_{\subset}^{(t)}$ that are updated in each \Mstep~of the \ac{EM} algorithm (see line~\ref{alg:EMschedule} in Alg.~\ref{alg:EMBP}). 
Finding an optimal \ac{EM} update schedule addresses challenges related to: optimizing the detection performance (\ref{itm:performance}), e.g., ``Is it favorable to perform multiple \ac{BP} iterations before updating the next parameter?'', and reducing  complexity (\ref{itm:complexity}), e.g., ``Given a limited number of \ac{EM} iterations~$T$ and  of total parameter updates~${K_\text{EM}:=\sum_{t=1}^T K^{(t)}}$, how does an optimal update schedule look like?''.
Learning a schedule is generally non-trivial as it corresponds to a discrete and combinatorial optimization.

For the purpose of integrating this problem into the discussed gradient-based end-to-end optimization framework, we convert the schedule learning task into the optimization of continuous weights, by leveraging the concept of momentum in the \ac{EM} parameter updates. Similarly to the discussed modification of the \ac{BP} message updates, we replace the original \ac{EM} parameter estimates $\hat{\theta}^{(t)}_k$ with momentum-based updates ${\beta_{\text{EM},k}^{(t)} \cdot \hat{{\theta}}^{(t)}_k} + {({1}-{\beta}_{\text{EM},k}^{(t)}) \cdot \hat{\theta}_k^{(t-1)}}$ where we have introduced the weights ${(\beta_{\text{EM},1}^{(t)},\ldots,\beta_{\text{EM},L+2}^{(t)})^{\rm T} =: \bm{\beta}_\text{EM}^{(t)} \in \mathbb{R}^{L+2}},\, {t=1,\ldots,T}$. %
In order to allow every possible schedule to be learned during the training procedure, we activate the computation of all parameter updates in the \Mstep\ by setting ${\hat{\bm{\theta}}_{\subset}^{(t)} = \hat{\bm{\theta}}^{(t)}}$ for all \ac{EM} iterations~$t$. The momentum weights are initialized with a serial schedule ${\bm{\beta}_\text{EM}^{(t)}} = {\bm{e}_{((t-1)~\text{mod}~L+2)+1}}$ at the beginning of the training phase, i.e., exactly one parameter is updated alternately per \ac{EM} iteration.
Then, the weights $\bm{\beta}_\text{EM}^{(t)}$ are tuned as part of the gradient-based optimization, converting the algorithm into a trainable discriminative  architecture~\cite{shlezinger2022discriminative}.

To learn an update schedule and  reduce the overall number of parameter updates $K_\text{EM} < T \cdot \left( L+2 \right)$, we add the $L_1$~regularizer that is known to encourage coefficients' sparsity~\cite[Chap.~3.1]{bishop_pattern_2006}
\begin{equation*}
    \mathcal{L}_\text{EM}(K') := \sum\limits_{\beta_{k} \in \mathcal{\beta}_{\text{min}}(K')} \lvert \beta_{k} \rvert
\end{equation*}
to the original loss function (\ac{MSE} or \ac{BMI}), where $\mathcal{\beta}_{\text{min}}(K')$ is the set of $K'$ scalar elements in the vectors $\bm{\beta}_\text{EM}^{(t)},\, t=1,\ldots,T$ with the smallest absolute value. Thereby, the regularization penalizes more than ${T\cdot\left( L+2 \right) - K'}$ parameter updates by enforcing $K'$ scalar elements of the vectors $\bm{\beta}_\text{EM}^{(t)},\, \forall t$ to be close to zero, i.e., they are effectively not being updated.
During training, we initialize ${K'=0}$ and gradually increase $K'$ until the target number of parameter updates ${K_\text{EM} = T\cdot\left( L+2 \right) - K'}$ is reached. After the training procedure, we implement the learned schedule by removing the $K'$ elements from $\hat{\bm{\theta}}_{\subset}^{(t)}$ that correspond to the respective elements in ${\mathcal{\beta}_{\text{min}}(K')}$. At this point,  EMBP  is reduced in complexity by reducing the number of required parameter update computations by~$K'$.
For the remaining parameter updates, we keep and apply the learned momentum weights $\bm{\beta}_\text{EM}^{(t)}$ as a simple and low-complexity generalization of the full \ac{EM} updates~(${\bm{\beta}_\text{EM}^{(t)}=\bm{1}}$).

\section{Experimental Study} \label{sec:experiments}
For the numerical experiments that are considered in this section, we will assume the transmission blocks to contain~${N=100}$ symbols from a \ac{BPSK} constellation. Additional simulation results for a \ac{QPSK} transmission are provided in Appendix~\ref{app:exp_channels}.
Unless explicitly mentioned otherwise, we apply ${S_\text{VAE}=10}$ iterations of the \ac{VAELE} with ${L_\text{LE}=2L}$ and a learning rate ${l_\text{VAE}=0.1}$ of the Adam optimizer. For the EMBP algorithm, we apply ${T=3\cdot \left( L+2 \right)}$ steps of the EMBP algorithm with a serial parameter update schedule, i.e., every parameter $\theta_\ell$ is updated $3$~times.

\subsection{Channel Estimation Analysis}
\begin{figure}
    \centering
    \begin{tikzpicture}[
        fill between/on layer={
            axis grid           %
        },
    ]
    \begin{axis}[
    width=\linewidth, %
    height=0.7\linewidth,
    align = left,
    grid=major, %
    grid style={gray!30}, %
    xlabel= ${\lVert \hat{\boldsymbol{h}}_\text{init} - \boldsymbol{h}\rVert^2}$,
    ylabel= ${\lVert \hat{\boldsymbol{h}} - \boldsymbol{h} \rVert^2}$,
	  scaled y ticks=false,
    ymode = log,
    xmode = log,
    ymin = 0.001,
    ymax = 20,
    xmin = 0.01,
    xmax = 100,
    enlarge x limits=false,
    enlarge y limits=false,
    line width=1pt,
	  legend style={font=\footnotesize, cells={align=left}, anchor=north west, at={(0.03,0.97)}},
    legend cell align={left},
    legend entries={${\hat{\boldsymbol{h}} = \hat{\boldsymbol{h}}_\text{init}}$, EMBP, VAE-LE},
    legend image code/.code={%
        \draw[dashed] (0cm,-0.05cm) -- (0.3cm,-0.05cm);
        \draw[solid] (0cm, 0.05cm) -- (0.3cm, 0.05cm);
    },
    legend style={xshift=0cm},
    ]
	\addplot[color=gray, line width=1pt] table[x={init mean}, y={init mean}, col sep=comma] {numerical_results/SE_noisyh_EsN010dB.csv};
	\addplot[color=EMBPDiraccolor, line width=1pt] table[x={init mean}, y={EMBP mean}, col sep=comma] {numerical_results/SE_noisyh_EsN010dB.csv};
	\addplot[color=VAEcolor, line width=1pt] table[x={init mean}, y={VAELE mean}, col sep=comma] {numerical_results/SE_noisyh_EsN010dB.csv};

	\addplot[color=gray, dashed, line width=1pt] table[x={init mean}, y={init median}, col sep=comma] {numerical_results/SE_noisyh_EsN010dB.csv};
    \addplot[color=EMBPDiraccolor, dashed, line width=1pt] table[x={init mean}, y={EMBP median}, col sep=comma] {numerical_results/SE_noisyh_EsN010dB.csv};
    \addplot[color=VAEcolor, dashed, line width=1pt] table[x={init mean}, y={VAELE median}, col sep=comma] {numerical_results/SE_noisyh_EsN010dB.csv};

    \addplot [EMBPDiraccolor, mark = o] coordinates {( 2.04, 0.072458)};
    \addplot [VAEcolor, mark = o] coordinates {( 2.04, 0.288214)};
 \addplot [EMBPDiraccolor, mark=star] coordinates {( 0.288214, 0.01191896 )};

\addplot[name path = init quantile25, color=gray, densely dotted] table[x={init mean}, y={init quantile25}, col sep=comma] {numerical_results/SE_noisyh_EsN010dB.csv};
 \addplot[name path = init quantile75, color=gray, densely dotted] table[x={init mean}, y={init quantile75}, col sep=comma] {numerical_results/SE_noisyh_EsN010dB.csv};
 \addplot [fill=gray!10] fill between[of=init quantile75 and init quantile25];
 \addplot[name path = EMBP quantile25, color=EMBPDiraccolor, densely dotted] table[x={init mean}, y={EMBP quantile25}, col sep=comma] {numerical_results/SE_noisyh_EsN010dB.csv};
 \addplot[name path = EMBP quantile75, color=EMBPDiraccolor, densely dotted] table[x={init mean}, y={EMBP quantile75}, col sep=comma] {numerical_results/SE_noisyh_EsN010dB.csv};
 \addplot [fill=EMBPDiraccolor!10] fill between[of=EMBP quantile75 and EMBP quantile25];
 \addplot[name path = VAELE quantile25, color=VAEcolor, densely dotted] table[x={init mean}, y={VAELE quantile25}, col sep=comma] {numerical_results/SE_noisyh_EsN010dB.csv};
 \addplot[name path = VAELE quantile75, color=VAEcolor, densely dotted] table[x={init mean}, y={VAELE quantile75}, col sep=comma] {numerical_results/SE_noisyh_EsN010dB.csv};
 \addplot [fill=VAEcolor!10] fill between[of=VAELE quantile75 and VAELE quantile25];
  \end{axis}
\end{tikzpicture}
    \caption{Squared error of the channel estimation~$\hat{\bm{h}}$ for the EMBP and VAE-LE algorithm at~${\mathsf{snr}=10}\,\text{dB}$ for $10^5$ random~channels with ${L=5}$, based on different initializations~$\hat{\bm{h}}_\text{init}$:
    (a)~${\hat{\bm{h}}_\text{init} = \bm{h} + \sqrt{\gamma} \bm{h}_w}$ with ${\bm{h}_w \sim \mathcal{CN}(0,\bm{I}_{L+1}), \gamma \in \mathbb{R}^+}$ (solid line: mean, dashed line: median, dotted lines: 25/75 percentiles), 
    (b)~${\hat{\bm{h}}_\text{init} = \bm{h}_{\bm{\delta}}}$ (circles: mean),
    (c)~initialization of the EMBP algorithm using the result~$\hat{\bm{h}}_\text{VAE}$ of the \ac{VAELE} with ${\hat{\bm{h}}_\text{init} = \bm{h}_{\bm{\delta}}}$ (star: mean). 
    The horizontal axis indicates the \ac{MSE} of the initialization ${\hat{\bm{h}}_\text{init}}$.}
    \label{fig:MSE_init}
\end{figure}
We start the numerical evaluation by analyzing the channel estimation performance of the proposed EMBP algorithm and the VAE-LE, for different initializations of the channel parameters~$\hat{\bm{h}}_\text{init}$. To this end, we sample $10^5$ random channels with  memory ${L=5}$ and corresponding channel observations $\bm{y}$ at ${\mathsf{snr}=10\,\text{dB}}$. 
Fig.~\ref{fig:MSE_init} evaluates the squared error~${\lVert \hat{\boldsymbol{h}} - \boldsymbol{h} \rVert^2}$ of the VAE-LE estimate $\hat{\bm{h}}_\text{VAE}$ and of the EMBP algorithm ($\hat{\bm{h}}_\text{EMBP}$) for three different initialization methods:
(a)~${\hat{\bm{h}}_\text{init} = \bm{h} + \sqrt{\gamma} \bm{h}_w}$ where ${\bm{h}_w \sim \mathcal{CN}(0,1)}$ is standard complex normally distributed, i.e., the initial estimate~$\hat{\bm{h}}_\text{init}$ is a noisy version of the true channel impulse response~$\bm{h}$ and  ${\gamma \in \mathbb{R}^+}$ indicates the variance of the noise. Note that this method is genie-aided due to its dependence on~$\bm{h}$; (b)~blind initialization with an impulse~${\hat{\bm{h}}_\text{init} = \bm{h}_{\bm{\delta}}}$; and (c)~initialization of the EMBP algorithm by using the estimation result of the \ac{VAELE} which is, in turn, initialized according to method~(b).

The results of method~(a) are given by the line plots in Fig.~\ref{fig:MSE_init}, where the horizontal axis indicates the squared initialization error ${\lVert \hat{\boldsymbol{h}}_\text{init} - \boldsymbol{h} \rVert^2 = \gamma \cdot (L+1)}$ averaged over all $10^5$ channel realizations. The vertical axis plots the quality of the channel estimation by the VAE-LE and EMBP algorithm.
Both algorithms have a monotonic improvement of the estimation quality with a decreasing initialization error. The \ac{MSE} of the \ac{VAELE} saturates to ${\lVert \hat{\boldsymbol{h}}_\text{VAE}-\boldsymbol{h} \rVert^2=0.24}$ for small initialization errors. We conjecture that this error floor is caused by the limited capabilities of the linear equalizer in the joint estimation and detection process of the \ac{VAELE}. 
In comparison, the \ac{MSE} of the EMBP algorithm saturates to a much smaller error floor.
We can observe that the mean and median substantially differ from each other which indicates the existence of outliers for the EMBP algorithm. %

Another noteworthy effect is the waterfall behavior of the median squared error for the EMBP algorithm. Nonlinear estimators are well-known to exhibit such a waterfall effect where the squared error increases abruptly below a certain threshold~\cite{van_trees_detection_2004}. This behavior is typically caused by very noisy observations which lead to a transition from local to global estimation errors in the \ac{ML} estimates~\cite{athley_threshold_2005}.

Comparing the results to initialization method~(b), it is worth mentioning that the latter can surpass the \ac{MSE} results of method~(a) with the same average initialization error for the EMBP algorithm. 
Finally, we analyze the proposed method~(c) where the \ac{VAELE} is leveraged to produce an initialization for the EMBP algorithm. 
The resulting estimates of the EMBP algorithm outperform
the estimation performance of method~(a) with the same mean squared initialization error~${\lVert \hat{\boldsymbol{h}}_\text{init} - \boldsymbol{h}\rVert^2=0.29}$. This confirms the \ac{VAELE} as a very well-suited method to initialize the EMBP algorithm.\\

We furthermore evaluate the squared error~${\lVert \hat{\boldsymbol{h}} - \boldsymbol{h} \rVert^2}$ for the considered methods as a function of the \ac{SNR} in Fig.~\ref{fig:MSE_vs_SNR_BPSK_L5}. In general, the estimation quality improves for higher~$\mathsf{snr}$. 
For the EMBP algorithm initialized according to method~(b) with~$\bm{h_\delta}$, we can again observe the deviation of the mean from the median especially for high~$\mathsf{snr}$.
This deviation can be significantly reduced by using the \ac{VAELE} for initialization: the \ac{MSE} of the EMBP algorithm with method~(c) is much closer to its respective median.
\begin{figure}[tb]
\centering
  \begin{tikzpicture}
    \pgfplotsset{
        legend image VAELE/.style={
            legend image code/.code={%
            \draw[solid, color=VAEcolor] (0cm,0.05cm) -- (0.3cm,0.05cm);
            \draw[dashed, color=VAEcolor] (0cm, -0.05cm) -- (0.3cm, -0.05cm);
            }
        },
    }
        \pgfplotsset{
        legend image EMBPb/.style={
            legend image code/.code={%
            \draw[solid, color=EMBPDiraccolor] (0cm,0.05cm) -- (0.3cm,0.05cm);
            \draw[dashed, color=EMBPDiraccolor] (0cm, -0.05cm) -- (0.3cm, -0.05cm);
            }
        },
    }
        \pgfplotsset{
        legend image EMBPc/.style={
            legend image code/.code={%
            \draw[solid,color=EMBPcolor] (0cm,0.05cm) -- (0.3cm,0.05cm);
            \draw[dashed,color=EMBPcolor] (0cm, -0.05cm) -- (0.3cm, -0.05cm);
            }
        },
    }
    \pgfplotsset{
        legend image EMNBPc/.style={
            legend image code/.code={%
            \draw[solid,color=EMNBPcolor] (0cm,0.05cm) -- (0.3cm,0.05cm);
            \draw[dashed,color=EMNBPcolor] (0cm, -0.05cm) -- (0.3cm, -0.05cm);
            }
        },
    }
    \pgfplotsset{
        legend image ML/.style={
            legend image code/.code={%
            \draw[color=black!100!white] (0cm,0.05cm) -- (0.3cm,0.05cm);
            \draw[color=black!60!white] (0cm, -0.05cm) -- (0.3cm, -0.05cm);
            }
        },
    }
    \pgfplotsset{
        legend image decdir/.style={
            legend image code/.code={%
            \draw[color=decdircolor!100!white] (0cm,0.05cm) -- (0.3cm,0.05cm);
            \draw[color=decdircolor!60!white] (0cm, -0.05cm) -- (0.3cm, -0.05cm);
            }
        },
    }
    \pgfplotsset{
legend image code/.code={
\draw[mark repeat=2,mark phase=2]
plot coordinates {
(0cm,0cm)
(0.15cm,0cm)        %
(0.3cm,0cm)         %
};%
}
}
    \begin{axis}[
    width=\linewidth, %
    height=0.7\linewidth,
    align = left,
    grid=major, %
    grid style={gray!30}, %
    xlabel= $\mathsf{snr}$ (dB),
    ylabel= ${\lVert \hat{\boldsymbol{h}} - \boldsymbol{h} \rVert^2}$,
	  scaled y ticks=false,
    ymode = log,
    ymin = 0.002,
    ymax = 1.0,
    xmin = -4,
    xmax = 12,
    enlarge x limits=false,
    enlarge y limits=false,
    line width=1pt,
	  legend style={font=\footnotesize, cells={align=left}, anchor=south west, at={(0.015,0.015)}},
    legend cell align={left},
    ]
    \addlegendimage{legend image VAELE}
    \addlegendentry{VAE-LE}
    \addlegendimage{legend image EMBPb}
    \addlegendentry{EMBP, init.~(b)}
    \addlegendimage{legend image EMBPc}
    \addlegendentry{EMBP, init.~(c)}
    \addlegendimage{legend image ML}
    \addlegendentry{ML, $10/20\%$ pilots}
    \addlegendimage{legend image decdir}
    \addlegendentry{DD-MAP, $10/20\%$ pilots}

    \addplot[color=black!100!white, line width=1pt] table[x={Es/N0 (dB)}, y={MSE h pilots}, col sep=comma] {numerical_results/decdir_BPSK_L5_10pilots.csv};
    \node at (axis cs:10,0.16) {\footnotesize ${10\%}$};
    \addplot[color=black!60!white, line width=1pt] table[x={Es/N0 (dB)}, y={MSE h pilots}, col sep=comma] {numerical_results/decdir_BPSK_L5_20pilots.csv};
    \node at (axis cs:11,0.028) {\footnotesize \color{black!60!white} ${20\%}$};

    \addplot[color=decdircolor!100!white, line width=1pt] table[x={Es/N0 (dB)}, y={MSE h decdirMAP}, col sep=comma] {numerical_results/decdir_BPSK_L5_10pilots.csv};
    \node at (axis cs:9.6,0.02) {\footnotesize \color{decdircolor!100!white} ${10\%}$};
    \addplot[color=decdircolor!60!white, line width=1pt] table[x={Es/N0 (dB)}, y={MSE h decdirMAP}, col sep=comma] {numerical_results/decdir_BPSK_L5_20pilots.csv};
    \node at (axis cs:9.6,0.005) {\footnotesize \color{decdircolor!60!white} ${20\%}$};
    
	\addplot[color=VAEcolor, line width=1pt] table[x={EsN0 dB}, y={VAE meanSE}, col sep=comma] {numerical_results/BER+MSE_over_EsN0_BPSK_L5.csv};
    \addplot[color=VAEcolor, line width=1pt, dashed, mark options={solid}] table[x={EsN0 dB}, y={VAE medianSE}, col sep=comma] {numerical_results/BER+MSE_over_EsN0_BPSK_L5.csv};
    \addplot[color=EMBPDiraccolor, line width=1pt] table[x={EsN0 dB}, y={EMBP(Dirac) meanSE}, col sep=comma] {numerical_results/BER+MSE_over_EsN0_BPSK_L5.csv};
    \addplot[color=EMBPDiraccolor, line width=1pt, dashed] table[x={EsN0 dB}, y={EMBP(Dirac) medianSE}, col sep=comma] {numerical_results/BER+MSE_over_EsN0_BPSK_L5.csv};
    \addplot[color=EMBPcolor, line width=1pt] table[x={EsN0 dB}, y={EMBP(VAE) meanSE}, col sep=comma] {numerical_results/BER+MSE_over_EsN0_BPSK_L5.csv};
    \addplot[color=EMBPcolor, line width=1pt, dashed] table[x={EsN0 dB}, y={EMBP(VAE) medianSE}, col sep=comma] {numerical_results/BER+MSE_over_EsN0_BPSK_L5.csv};
  \end{axis}
\end{tikzpicture}
    \caption{\ac{MSE} versus $\mathsf{snr}$ for $10^7$ random channels with ${L=5}$ for various algorithms and initialization methods (solid line: mean, dashed line: median).} 
  \label{fig:MSE_vs_SNR_BPSK_L5}
\end{figure}%

We compare the results of the blind estimation methods to a pilot-based channel estimation. To this end, we fix a portion of the $N$ information symbols in the transmission block to pseudo-random pilot symbols that are known at the receiver. Based on these pilot symbols, the receiver performs an \ac{ML} estimation of~$\bm{h}$.\footnote{For the pilot-aided baseline methods, we only estimate the channel impulse response~$\bm{h}$ and assume perfect knowledge of $\sigma^2$, as opposed to the blind schemes.} Fig.~\ref{fig:MSE_vs_SNR_BPSK_L5} shows the performance of the consequent \ac{ML} estimation based on ${N\cdot10\%}$ and ${N\cdot20\%}$ pilot symbols (gray lines). 
Further, we consider a \ac{DD} scheme which uses not only the pilot symbols for channel estimation but also incorporates the remaining received symbols in~$\bm{y}$ to acquire more precise \ac{CSI}.
To this end, we perform \ac{MAP} detection using the already obtained pilot-based channel estimates. A refined \ac{ML} channel estimation is then performed by incorporating the entire data block based on the pilot symbols together with a hard decision of the \ac{MAP} detection for the remaining symbols.
The results are shown in Fig.~\ref{fig:MSE_vs_SNR_BPSK_L5} (red lines) and confirm the good estimation performance of the blind EMBP scheme with initialization method~(c), which consistently outperforms the \ac{DD}-\ac{MAP} baseline with $10\%$ pilots and competes with the variant that uses $20\%$ pilots.

\subsection{Symbol Detection Evaluation}
\begin{figure}
  \centering
      \begin{tikzpicture}
      \pgfplotsset{
        legend image solid/.style={
            legend image code/.code={%
            \draw[color=black!100!white] (0cm,0.05cm) -- (0.3cm,0.05cm);
            \draw[color=black!60!white] (0cm, -0.05cm) -- (0.3cm, -0.05cm);
            }
        },
    }
    \pgfplotsset{
        legend image dotted/.style={
            legend image code/.code={%
            \draw[dashdotted, color=black!100!white] (0cm,0.05cm) -- (0.3cm,0.05cm);
            \draw[dashdotted, color=black!60!white] (0cm, -0.05cm) -- (0.3cm, -0.05cm);
            }
        },
    }
    \pgfplotsset{
        legend image decdir/.style={
            legend image code/.code={%
            \draw[dashdotted, color=decdircolor!100!white] (0cm,0.05cm) -- (0.3cm,0.05cm);
            \draw[dashdotted, color=decdircolor!60!white] (0cm, -0.05cm) -- (0.3cm, -0.05cm);
            }
        },
    }
    \pgfplotsset{
legend image code/.code={
\draw[mark repeat=2,mark phase=2]
plot coordinates {
(0cm,0cm)
(0.15cm,0cm)        %
(0.3cm,0cm)         %
};%
}
}
    \begin{axis}[
    width=\linewidth, %
    height=0.8\linewidth,
    align = left,
    grid=major, %
    grid style={gray!30}, %
    xlabel= $\mathsf{snr}$ (dB),
    ylabel= BER,
	  scaled y ticks=false,
    ymode = log,
    ymin = 0.0003,
    ymax = 0.5,
    xmin = -4,
    xmax = 12,
    enlarge x limits=false,
    enlarge y limits=false,
    line width=1pt,
	  legend style={font=\footnotesize, cells={align=left}, anchor=south west, at={(0.015,0.015)}},
    legend cell align={left},
	  smooth,
    ]
    \addplot[color=VAEcolor, line width=1pt] table[x={EsN0 dB}, y={VAE BER}, col sep=comma] {numerical_results/BER+MSE_over_EsN0_BPSK_L5.csv};
    \addlegendentry{VAE-LE}
    \addplot[color=EMBPcolor, line width=1pt] table[x={EsN0 dB}, y={EMBP(VAE) BER}, col sep=comma] {numerical_results/BER+MSE_over_EsN0_BPSK_L5.csv};
    \addlegendentry{EMBP}
    \addplot[color=EMNBPcolor, line width=1pt] table[x={EsN0 dB}, y={EMNBP(VAE) BER}, col sep=comma] {numerical_results/BER+MSE_over_EsN0_BPSK_L5.csv};
    \addlegendentry{EMBP$^\star$, $\beta_\text{BP}^\star$}
    \addplot[color=BPcolor, line width=1pt, densely dotted] table[x={EsN0 dB}, y={BP(genie) BER}, col sep=comma] {numerical_results/BER+MSE_over_EsN0_BPSK_L5.csv};
    \addlegendentry{BP, $\bm{h}$ (coherent)}
    \addplot[color=black!30!white, line width=1pt, densely dotted] table[x={EsN0 dB}, y={MAP BER}, col sep=comma] {numerical_results/MAP_BER_over_EsN0_BPSK_L5.csv};
    \addlegendentry{MAP, $\bm{h}$ (coherent)}
    \addlegendimage{legend image dotted}
    \addlegendentry{MAP, $\hat{\bm{h}}_\text{ML}$, \\ $10/20\%$ pilots}
    \addlegendimage{legend image decdir}
    \addlegendentry{MAP, $\hat{\bm{h}}_\text{DD-MAP}$, \\ $10/20\%$ pilots}
    \addplot[color=black!100!white, line width=1pt, dashdotted] table[x={Es/N0 (dB)}, y={BER MAP pilots}, col sep=comma] {numerical_results/decdir_BPSK_L5_10pilots.csv};
    \addplot[color=black!60!white, line width=1pt, dashdotted] table[x={Es/N0 (dB)}, y={BER MAP pilots}, col sep=comma] {numerical_results/decdir_BPSK_L5_20pilots.csv};
    \node at (axis cs:11,0.02) {\footnotesize \color{black!100!white} ${10\%}$};
    \node at (axis cs:10,0.001) {\footnotesize \color{black!60!white} ${20\%}$};
     
    \addplot[color=decdircolor!100!white, line width=1pt, dashdotted] table[x={Es/N0 (dB)}, y={BER MAP decdirMAP}, col sep=comma] {numerical_results/decdir_BPSK_L5_10pilots.csv};
    \addplot[color=decdircolor!60!white, line width=1pt, dashdotted] table[x={Es/N0 (dB)}, y={BER MAP decdirMAP}, col sep=comma] {numerical_results/decdir_BPSK_L5_20pilots.csv};
    \node at (axis cs:9.1,0.013) {\footnotesize \color{decdircolor!100!white} ${10\%}$};
    \node at (axis cs:8.3,0.001) {\footnotesize \color{decdircolor!60!white} ${20\%}$};
  \end{axis}
\end{tikzpicture}
\caption{\ac{BER} over $\mathsf{snr}$ for various detection schemes, averaged over $10^7$ random channels with ${L=5}$.}
\label{fig:BER_vs_SNR_BPSK_L5}
\end{figure}
Based on the same data that was sampled for the results in Fig.~\ref{fig:MSE_vs_SNR_BPSK_L5}, we evaluate the detection performance of the considered joint estimation and detection schemes in terms of the \ac{BER} in Fig.~\ref{fig:BER_vs_SNR_BPSK_L5}.
For low ${\mathsf{snr}}$, the \ac{BER} performance of the \ac{VAELE} is relatively close to the optimal coherent \ac{MAP} performance.
This performance gap increases significantly for higher~$\mathsf{snr}$ where the \ac{BER} of the VAE-LE only improves marginally.
The sampled channels can introduce severe \ac{ISI} where linear equalizers like the \ac{FIR} filter of the \ac{VAELE}\footnote{For channels with less severe \ac{ISI}, the \ac{VAELE} typically performs significantly better, as demonstrated in~\cite{lauinger_blind_2022}.} perform poorly~\cite{proakis_digital_2007}.
The EMBP algorithm performs significantly better. For low $\mathsf{snr}$, it can be compared to \ac{MAP} detection performance based on the \ac{DD}-\ac{MAP} estimation baseline with $20\%$ pilots.
In the high \ac{SNR} regime, the EMBP algorithm runs into an error floor.
This effect is mainly caused by the suboptimality of the \ac{BP} detector due to many short cycles in the underlying factor graph, which can cause a non-convergent behavior of the \ac{BP} algorithm. Especially for high ${\mathsf{snr}}$, the dynamic range of the factor nodes~\eqref{eq:f_fn} and~\eqref{eq:I_fn} is further increased~\cite{schmid_low-complexity_2022} and the suboptimality of \ac{BP} becomes the dominant error source.
To reduce this error floor, we can leverage the momentum-based \ac{BP} message updates as introduced in Sec.~\ref{subsubsec:momentumBP}. We optimize the weights ${\beta_\text{BP}^{(t)}, t=1,\ldots,L+2}$ for generic channels, i.e., each sample in the training data set $\mathcal{D}$ consists of an independent channel realization and the ${\mathsf{snr}}$ is uniformly sampled in $[0,12]\,\text{dB}$. This generic   training can be performed offline and prior to evaluation.
We apply batch gradient descent optimization with $200$ batches each containing $1000$ transmission blocks. 
As shown in Fig.~\ref{fig:BER_vs_SNR_BPSK_L5}, the refined EMBP$^\star$ algorithm with optimized momentum weights~$\beta_\text{BP}^\star$ can effectively reduce the error floor\footnote{The lowered but remaining error floor of the EMBP$^\star$ algorithm in Fig.~\ref{fig:BER_vs_SNR_BPSK_L5} vanishes by the application of \ac{FEC}, as demonstrated in Appendix~\ref{app:fec}.} and thereby reduces the gap to coherent \ac{MAP} detection significantly for a target ${\text{BER}=10^{-2}}$.

A surprising observation in Fig.~\ref{fig:BER_vs_SNR_BPSK_L5} is that the blind EMBP algorithm clearly outperforms the coherent \ac{BP} detector for ${\mathsf{snr} > 3\,\text{dB}}$. Although the EMBP algorithm was demonstrated to produce good channel estimates in Fig.~\ref{fig:MSE_vs_SNR_BPSK_L5}, it seems counterintuitive that a blind detector is able to outperform its coherent counterpart which has full access to the \ac{CSI}.
While this intuition is valid for optimal \ac{MAP} detection, it is not given for suboptimal detectors such as \ac{BP} on cyclic factor graphs, whose suboptimality directly and heavily depends on the characteristics of the underlying channel assumption.
To exemplify this phenomenon, we take a closer look at the transmission of $10^4$ random information blocks over one specific channel with impulse response~$\bm{h}_0 = ({0.3-0.3\text{j}},\allowbreak {0.6-0.1\text{j}},\allowbreak {0.6-0.3\text{j}})^{\rm T}$ and ${\mathsf{snr} = 10\,\text{dB}}$.
The coherent \ac{BP} detector (${\text{BER}=0.21}$) is significantly outperformed by the EMBP algorithm with a ${\text{BER}=0.051}$. The reason for this improved detection performance becomes obvious if we evaluate the mean of the channel estimates~$\hat{\bm{h}}_{0,\text{EMBP}} = ({0.22-0.26\text{j}},\allowbreak {0.51-0.063\text{j}},\allowbreak {0.51-0.26\text{j}})^{\rm T}$: the channel taps of the EMBP estimate are consistently reduced in their amplitude compared to the true channel~$\bm{h}_0$. This directly influences the dynamic range of the factor nodes~\eqref{eq:f_fn} and~\eqref{eq:I_fn}, known to notably affect the convergence of \ac{BP} on factor graphs with cycles~\cite{ihler_loopy_2005}.

\begin{figure}
    \centering
    \begin{tikzpicture}
        \pgfplotsset{
    legend image code/.code={
    \draw[mark repeat=2,mark phase=2]
    plot coordinates {
    (0cm,0cm)
    (0.15cm,0cm)        %
    (0.3cm,0cm)         %
    };%
    }
    }
\begin{groupplot}[
        group style={
            group size=1 by 2,
            x descriptions at=edge bottom,
            vertical sep=3pt,
        },
    ]
    
\nextgroupplot[
    ymode=log,
    width=\linewidth, %
    height= 0.4\linewidth,
    align = left,
    grid=major, %
    grid style={gray!30}, %
    ylabel= {BER},
    ymin = 0.001,
    ymax = 0.5,
    xmin = 0.0,
    xmax = 1.4,
    enlarge x limits=false,
    enlarge y limits=false,
    line width=1pt,
    legend style={font=\footnotesize, cells={align=left}, anchor=south west, at={(0.03,0.04)}},
    legend cell align={left},
    ]
    \addplot[thick, samples=2, dashed, gray!50!white, name path=ML] coordinates {(1.0,0.0001)(1.0,0.5)};
    \addplot[thick, samples=2, dashed, EMBPcolor!50!white, name path=ML] coordinates {(0.61,0.02)(0.61,0.0001)};
    \addplot[color=EMBPcolor, line width=1pt] table[x={alpha}, y={BER(BP)}, col sep=comma] {numerical_results/ELBO_vs_alpha.csv};
    \addplot[color=gray, line width=1pt, densely dotted] table[x={alpha}, y={BER(MAP)}, col sep=comma] {numerical_results/ELBO_vs_alpha.csv};

    \nextgroupplot[
    width=\linewidth, %
    height=0.5\linewidth,
    align = left,
    grid=major, %
    grid style={gray!30}, %
    xlabel= $\alpha$,
    ylabel= {$\mathcal{L}(Q,\bm{\theta}) \cdot 10^{-2}$},
    ymin = -2000,
    ymax = 0,
    ytick={-2000,-1500, -1000, -500},
    yticklabels={$-20$,$-15$, $-10$, $-5$},
    xmin = 0.0,
    xmax = 1.4,
    enlarge x limits=false,
    enlarge y limits=false,
    line width=1pt,
    legend style={font=\footnotesize, cells={align=left}, anchor=south west, at={(0.03,0.05)}},
    legend cell align={left},
    ]
    \addplot[draw=none, color=gray, line width=1pt, densely dotted] table[x={alpha}, y={log-likelihood}, col sep=comma] {numerical_results/ELBO_vs_alpha.csv};
    \addlegendentry{$Q_\text{APP}=P=P(\bm{c}|\bm{y},{\bm{h}=\alpha \bm{h}_0})$}
    \addplot[draw=none, color=EMBPcolor, line width=1pt] table[x={alpha}, y={ELBO(BP)}, col sep=comma] {numerical_results/ELBO_vs_alpha.csv};
    \addlegendentry{$Q_\text{BP}=\text{BP}(\bm{y},{\bm{h}=\alpha \bm{h}_0})$} %
    \addplot[thick, samples=2, dashed, gray!50!white, name path=ML] coordinates {(1,-82.7)(1,-2000)};
    \addplot[thick, samples=2, dashed, EMBPcolor!50!white, name path=ML] coordinates {(0.66,-212.6)(0.66,-2000)};
    \addplot[color=EMBPcolor, line width=1pt] table[x={alpha}, y={ELBO(BP)}, col sep=comma] {numerical_results/ELBO_vs_alpha.csv};
    \addplot[color=gray, line width=1pt, densely dotted] table[x={alpha}, y={log-likelihood}, col sep=comma] {numerical_results/ELBO_vs_alpha.csv};
    \draw [latex-latex, magenta, thick, name path=KL] (1.027,-1504) -- node[midway, right] {\small $D_\text{KL}(Q_\text{BP} \Vert P)$} (1.027,-90.8);
    \end{groupplot}
    \end{tikzpicture}
    \caption{\ac{BER} and \ac{ELBO}~${\mathcal{L}(Q,\bm{\theta})}$ of the exact \ac{APP} distribution $Q_\text{APP}$ and the approximate distribution $Q_\text{BP}$ for different channel assumptions~${\bm{h}=\alpha \bm{h}_0}$.}
    \label{fig:ELBO_vs_alpha}
\end{figure}
But why does the EMBP algorithm converge to such a ``surrogate'' channel~$\hat{\bm{h}}_{0,\text{EMBP}}$, despite the fact that it is based on \ac{ML} estimation? To answer this question, we recall that the \ac{EM} algorithm simplifies \ac{ML} estimation to the maximization of the \ac{ELBO}~$\mathcal{L}(Q,\bm{\theta})$ which is only equivalent to maximizing the likelihood function if the \Estep\ is using \emph{exact} \ac{APP} distributions $P(\bm{c}|\bm{y},\hat{\bm{\theta}}^{(t-1)})$. Fig.~\ref{fig:ELBO_vs_alpha} visualizes $\mathcal{L}(Q,\bm{\theta})$ both for the true posterior distribution~${Q_\text{APP}=P(\bm{c}|\bm{y},\bm{\theta})}$ and for the approximative \ac{BP} solution $Q_\text{BP}=\text{BP}(\bm{y},\bm{\theta})$ over a one-dimensional subspace ${\{\alpha \bm{h}_0: \, \alpha \in \mathbb{R}^+\}}$ of the complete estimation space of $\bm{h}$ at ${\mathsf{snr}=10\,\text{dB}}$.
For the \ac{APP} distribution, ${D_\text{KL}(Q_\text{APP}||P)=0}$ holds and thus ${\mathcal{L}(Q_\text{APP},\bm{\theta})}={\log p(\bm{y}|\bm{\theta})}$. The maximum of the \ac{ELBO}~$\mathcal{L}(Q_\text{APP},\bm{\theta})$ is therefore at ${\alpha=1}$, i.e., the \ac{EM} algorithm converges to the true channel ${\bm{h}=\bm{h}_0}$. 
In contrast, the \ac{KL} divergence between $Q_\text{BP}$ and $P$ is non-vanishing for ${\alpha>0.6}$ and rapidly grows with increasing $\alpha$.
This degrading accuracy of the \ac{BP} algorithm for large~$\alpha$ can also be observed when evaluating the \ac{BER} of the \ac{BP}-based detection which has its minimum at ${\alpha = 0.61}$.
The maximum of the \ac{ELBO} for \ac{BP} consequently lies at ${\alpha = 0.66}$ between the \ac{ML} solution (${\alpha = 1.0}$) and a small \ac{KL} divergence (${\alpha < 0.6}$). This spot corresponds to the ``surrogate'' channel to which the EMBP algorithm converges by maximizing the \ac{ELBO}~$\mathcal{L}(Q_\text{BP},\bm{\theta})$\footnote{Note that the damping of the channel taps with a \emph{scalar} factor $\alpha \in \mathbb{R}^+$ in Fig.~\ref{fig:ELBO_vs_alpha} is only for illustrative purposes and does not represent the true maximum within the complete estimation space of $\bm{\theta}$.}.
Thereby,  EMBP   finds a channel representation that is better suited for \ac{BP}-based symbol detection compared to the ground truth channel.\\

The channels considered so far follow a uniformly distributed \ac{pdp} as defined in Sec.~\ref{subsec:system_model}. However, many realistic communication channels follow a \ac{pdp} that is non-uniformly distributed, e.g., exponentially decaying.
The behavior of the proposed EMBP algorithm on this alternative channel characteristic behaves qualitatively similar to the previously discussed results. Detailed evaluations for this case are provided in Appendix~\ref{app:exp_channels}, which also contains results for a \ac{QPSK} transmission.

\begin{figure}
    \centering
    \begin{tikzpicture}
    \begin{axis}[
    ymode=log,
    width=\linewidth, %
    height= 0.5\linewidth,
    align = left,
    grid=major, %
    grid style={gray!30}, %
    xlabel= ${S_\text{VAE}-s}$ \hspace{2.5cm} $t$ \hspace{1cm},
    ylabel= ${\lVert \hat{\boldsymbol{h}} - \boldsymbol{h} \rVert^2}$,
    ymin = 0.007,
    ymax = 1.8,
    xmin = -9.5,
    xmax = 12.5,
    xtick = {-9,-6,-3,0,1,3,5,7,9,11,13},
    xticklabels = {$9$,$6$,$3$,$0$, $1$, $3$, $5$, $7$, $9$, $11$, $13$},
    enlarge x limits=false,
    enlarge y limits=false,
    line width=1pt,
    legend style={font=\footnotesize, cells={align=left}, anchor=south west, at={(0.0, 1.04)}},
    legend cell align={left},
    ]
    \addplot[color=VAEcolor, mark=square*, line width=0.5pt, only marks] table[x={iters}, y={VAE SE h}, col sep=comma] {numerical_results/eval_over_iters.csv};
    \addlegendentry{VAE-LE ${(S_\text{VAE}=10},$ ${l_\text{VAE} = 0.1})$}
    \addplot[draw=none, color=EMBPcolor, mark=square*, line width=0.5pt, only marks] coordinates {(-12,0)(-12,1)};
    \addlegendentry{EMBP ${(T=12}$, serial schedule$)$}
    \addplot[draw=none, color=EMBPcolor, mark=square, line width=0.5pt, only marks] coordinates {(-12,0)(-12,1)};
    \addlegendentry{EMBP ${(T=12}$, parallel schedule$)$}
    \addplot[color=EMNBPcolor, mark=*, line width=0.9pt, only marks, mark options={scale=0.85}] table[x={iters}, y={VAEopt3 SE h}, col sep=comma] {numerical_results/eval_over_iters.csv};
    \addlegendentry{VAE-LE ${(S_\text{VAE}=3)}$ $+$ EMBP$^\star$ ${(T=3)}$}
    \addplot[color=EMNBPcolor, mark=o, line width=0.5pt, only marks, mark options={scale=0.85}] table[x={iters}, y={VAEopt6 SE h}, col sep=comma] {numerical_results/eval_over_iters.csv};
    \addlegendentry{VAE-LE ${(S_\text{VAE}=3)}$ $+$ EMBP$^\star$ ${(T=6)}$}
    \addplot[color=EMBPcolor, mark=square*, line width=0.5pt, only marks] table[x={iters}, y={EMBPserial SE h}, col sep=comma] {numerical_results/eval_over_iters.csv};
    \addplot[color=EMBPcolor, mark=square, line width=0.5pt, only marks] table[x={iters}, y={EMBPparallel SE h}, col sep=comma] {numerical_results/eval_over_iters.csv};
    \addplot[color=EMNBPcolor, mark=o, line width=0.5pt, only marks, mark options={scale=0.85}] table[x={iters}, y={EMBPopt6 SE h}, col sep=comma] {numerical_results/eval_over_iters.csv};
    \addplot[color=EMNBPcolor, mark=*, line width=0.9pt, only marks, mark options={scale=0.85}] table[x={iters}, y={EMBPopt3 SE h}, col sep=comma] {numerical_results/eval_over_iters.csv};

    \addplot[color=VAEcolor, mark=square*, line width=0.5pt, only marks] table[x={iters}, y={VAE SE h}, col sep=comma] {numerical_results/eval_over_iters.csv};
    \addplot[color=EMNBPcolor, mark=*, line width=0.5pt, only marks, mark options={scale=0.85}] table[x={iters}, y={VAEopt6 SE h}, col sep=comma] {numerical_results/eval_over_iters.csv};
    \addplot[color=EMNBPcolor, mark=o, line width=0.9pt, only marks, mark options={scale=0.85}] table[x={iters}, y={VAEopt3 SE h}, col sep=comma] {numerical_results/eval_over_iters.csv};

    \node at (axis cs:-4.5,1.0) {\footnotesize VAE-LE};
    \draw [latex-, gray!70!white, thick, name path=VAELE left] (-9.5,1.0) -- (-6.2,1.0);
    \draw [-latex, gray!70!white, thick, name path=VAELE right] (-2.7,1.0) -- (0.5,1.0);
    \node at (axis cs:6.0,1.0) {\footnotesize EMBP};
    \draw [latex-, gray!70!white, thick, name path=EMBP left] (0.5,1.0) -- (4.5,1.0);
    \draw [-latex, gray!70!white, thick, name path=EMBP right] (7.5,1.0) -- (12.5,1.0);
    \end{axis}
    \end{tikzpicture}
    \caption{Mean squared estimation error ${\lVert \hat{\boldsymbol{h}} - \boldsymbol{h} \rVert^2}$ after each iteration of the \ac{VAELE} and the EMBP algorithm with different \ac{EM} parameter update schedules. The randomly sampled channels have ${\mathsf{snr}=10\,\text{dB}}$ and memory ${L=5}$, i.e., $7$ parameters to estimate (including $\sigma^2$).}
    \label{fig:SE_over_iters}
\end{figure}
\subsection{Complexity Reduction using Model-based Deep Learning}
We study the capability of the proposed model-based deep learning methods to reduce the computational complexity of the EMBP algorithm. To this end, we consider a transmission over randomly sampled channels with memory ${L=5}$ and optimize the EMBP algorithm towards minimum \ac{MSE} of the channel estimates~$\hat{\bm{h}}$ while restraining the algorithm's complexity. More specifically, we reduce the number of \ac{VAELE} iterations for the initialization from ${S_\text{VAE}=10}$ to ${S_\text{VAE}=3}$ and limit the \ac{EM} iterations to ${T=6}$ or ${T=3}$, respectively. Based on this restriction ${T < L+2=7}$, the EMBP algorithm would not be able to update each parameter (including $\sigma^2$) at least once if it followed a serial update schedule where only one parameter is updated per \ac{EM} iteration. We apply the gradient-based end-to-end training method, discussed in Sec.~\ref{sec:optimization}, to find a more effective \ac{EM} update schedule for this setting.
Besides the parameters ${\bm{\beta}_\text{EM}^{(t)}}, \, t=1,\ldots,T$ that define the \ac{EM} update schedule, we also include the momentum \ac{BP} weights ${\beta_\text{BP}^{(t)}}$ in the end-to-end optimization. 
We use $2500$ batches for the offline training and each batch contains $1000$ random channel realizations. For the schedule learning, we fix the maximum number of parameter updates to ${K_\text{EM}=24}$. 

To mitigate the effect of the reduced number of VAE-LE iterations~$S_\text{VAE}$, we additionally apply a hyperparameter search to obtain the optimized learning rates ${(l_\text{VAE}^{(1)}, l_\text{VAE}^{(2)}, l_\text{VAE}^{(3)})=(0.1,0.16,0.3)}$ for each \ac{VAELE} iteration~${s=1,2,3}$.
The left side in Fig.~\ref{fig:SE_over_iters} shows the squared estimation error after each \ac{VAELE} iteration compared to a baseline with ${S_\text{VAE}=10}$ \ac{VAELE} steps using a constant learning rate ${l_\text{VAE}=0.1}$. The results are averaged over $10^5$ random channels with ${\mathsf{snr}=10\,\text{dB}}$. We can observe that the \ac{VAELE} with optimized learning rates reduces the estimation error faster and achieves the same $\text{MSE}=0.36$ after only $3$ iterations for which the baseline \ac{VAELE} requires 6 steps.
\begin{figure}
    \centering
\begin{tikzpicture}
\def\deltax{0.85}
\def\deltay{0.56}
  \foreach \y [count=\n] in {
{0.0,0.0,0.0,0.63,0.0,0.0,0.0},
{0.76,0.84,0.0,0.0,0.82,0.78,0.0},
{0.70,0.64,1.60,0.0,0.80,0.83,0.0},
{0.77,0.96,0.0,1.26,0.0,0.97,0.93},
{0.64,0.0,0.84,0.0,0.84,0.0,0.0},
{1.15,1.18,1.39,1.34,1.19,1.16,0.0},
    } {
      \ifnum\n<7
        \node[minimum size=1em] at (\n*\deltax, 0) {\small $\hat{h}_{\n}$};
      \fi
      \foreach \x [count=\m] in \y {
        \node[fill=EMNBPcolor!\fpeval{120.0*\x-50.0}!white, minimum size=5mm, text=black] at (\m*\deltax,-\n*\deltay) {\small \x};
      }
    }
 \node[minimum size=1em] at (7*\deltax, 0) {$\hat{\sigma}^2$};
  \foreach \a [count=\i] in {1,2,3,4,5,6} {
    \node[minimum size=6mm] at (0,-\i*\deltay) {\small \a};
  }
  \node[rotate=90] at (-0.7*\deltax, -3.5*\deltay) {EMBP$^\star$ iteration~$t$};
  \node[] at (4*\deltax, 1.0*\deltay) {$\bm{\beta}_\text{EM}^{(t)}$};
\end{tikzpicture}
    \caption{Optimized \ac{EM} schedule of the EMBP$^\star$ algorithm based on the momentum weights $\bm{\beta}_\text{EM}^{(t)}$ for the complexity constraints ${T=6}$ and ${K_\text{EM}=24}$.}
    \label{fig:schedule}
\end{figure}

Based on the \ac{VAELE} initialization, the right side in Fig.~\ref{fig:SE_over_iters} shows the estimation performance of the EMBP algorithm over the \ac{EM} iterations~$t$.
Using a serial schedule, the \ac{MSE} monotonically decreases.
The stagnation at ${t=7}$ is due to the parameter estimate $\hat{\sigma}^2$ not being \emph{directly} included in the evaluation of the squared error ${\lVert \hat{\boldsymbol{h}} - \boldsymbol{h} \rVert^2}$.
The second baseline in Fig.~\ref{fig:SE_over_iters} uses a parallel schedule for   EMBP, i.e., all parameters are simultaneously updated in every \ac{EM} step~$t$. As a result, the \ac{MSE} decreases significantly after the first \ac{EM} iteration but worsens after ${t=2}$ due to an unstable behavior of the fully parallel updates.
The EMBP$^\star$ algorithm with optimized update schedules shows a fast yet stable convergence and the estimation results are comparable to those of the serial \ac{EM} schedule but with less iterations, i.e., with a reduced computational complexity.

The momentum weights~$\bm{\beta}_\text{EM}^{(t)}$ which define the corresponding learned schedule for ${T=6}$ are visualized in Fig.~\ref{fig:schedule}. 
In the first EMBP iteration ${t=1}$ there is only one parameter $\hat{h}_4$ updated with small weight, i.e., the algorithm basically performs two \ac{BP} iterations before starting to update the channel estimate. In the following EMBP iterations ${t=2,\ldots,5}$, multiple parameters are updated simultaneously. Only in the last iteration ${t=6}$, all channel taps are updated in parallel. 

\subsection{Comparison to Data-aided Receivers}
At last, we compare the blind EMBP algorithm to state-of-the-art data-aided deep receivers. As a model-agnostic approach, we consider a \ac{RNN} with window size ${L_\text{h}+1}$ followed by a linear layer and the softmax function. Further, we use the ViterbiNet detector~\cite{shlezinger_viterbinet_2020} as a model-based deep learning method.
We consider the transmission over a time-invariant channel with a real-valued and exponentially decaying impulse response~${\bm{h}_\text{e} = (0.802, 0.487, 0.295, 0.179)^{\rm T}}$. 
Both data-aided detectors are provided with one transmission block, i.e., ${N=100}$ \ac{BPSK} symbols, of labeled training data based on which they perform $500$ epochs of gradient descent optimization to adapt to the channel $\bm{h}_\text{e}$.  For a more involved elaboration on both considered algorithms, their respective training, and an open-source implementation, we refer to~\cite{raviv_adaptive_2023}.

\begin{figure}
    \centering
    \begin{tikzpicture}
    \pgfplotsset{
legend image code/.code={
\draw[mark repeat=2,mark phase=2]
plot coordinates {
(0cm,0cm)
(0.15cm,0cm)        %
(0.3cm,0cm)         %
};%
}
}
    \begin{axis}[
    width=\linewidth, %
    height=0.7\linewidth,
    align = left,
    grid=major, %
    grid style={gray!30}, %
    xlabel= $\mathsf{snr}$ (dB),
    ylabel= BER,
    scaled y ticks=false,
    ymode = log,
    ymin = 0.001,
    ymax = 0.5,
    xmin = -4,
    xmax = 12,
    enlarge x limits=false,
    enlarge y limits=false,
    line width=1pt,
	  legend style={font=\footnotesize, cells={align=left}, anchor=south west, at={(0.03,0.03)}},
    legend cell align={left},
	  smooth,
    ]
    \addplot[color=EMBPcolor, line width=1pt] table[x={Es/N0 (dB)}, y={EMBP BER}, col sep=comma] {numerical_results/BER_over_EsN0_exp4channel.csv};
    \addlegendentry{EMBP}
    \addplot[color=EMNBPcolor, line width=1pt] table[x={Es/N0 (dB)}, y={EMBP*L3 BER}, col sep=comma] {numerical_results/BER_over_EsN0_exp4channel.csv};
    \addlegendentry{EMBP$^\star$}
    \addplot[color=BPcolor, line width=1pt, densely dotted] table[x={Es/N0 (dB)}, y={genie BP BER}, col sep=comma] {numerical_results/BER_over_EsN0_exp4channel.csv};
    \addlegendentry{BP (coherent)}
    \addplot[color=black!30!white, line width=1pt, densely dotted] table[x={Es/N0 (dB)}, y={genie MAP BER}, col sep=comma] {numerical_results/BER_over_EsN0_exp4channel.csv};
    \addlegendentry{ MAP (coherent)}
    \addplot[color=RNNcolor, line width=1pt, dashdotted] table[x={Es/N0 (dB)}, y={RNN 100 pilots BER}, col sep=comma] {numerical_results/BER_over_EsN0_exp4channel.csv};
    \addlegendentry{RNN, ${100}$ pilots}
    \addplot[color=ViterbiNetcolor, line width=1pt, dashdotted] table[x={Es/N0 (dB)}, y={ViterbiNet 100 pilots BER}, col sep=comma] {numerical_results/BER_over_EsN0_exp4channel.csv};
    \addlegendentry{ViterbiNet, ${100}$ pilots}
  \end{axis}
\end{tikzpicture}
    \caption{Comparison of the blind EMBP algorithm to data-aided deep receivers (RNN and ViterbiNet) with respect to the \ac{BER} over $\mathsf{snr}$ for the channel~$\bm{h}_\text{e}$, averaged over $10^5$ random transmission blocks.}
    \label{fig:BER_vs_SNR_BPSK_L5_exp4}
\end{figure}
Fig.~\ref{fig:BER_vs_SNR_BPSK_L5_exp4} reports the \ac{BER} versus $\mathsf{snr}$, averaged over $10^5$ transmission blocks. Both the \ac{RNN}-based detector and the ViterbiNet algorithm were trained specifically for each $\mathsf{snr}$ value and the evaluation results are averaged over $50$ independent training runs with varying initialization.
In contrast, the EMBP detector is adaptive to the channel by design without the requirement of channel-specific training. This also holds for the EMBP$^\star$ algorithm which was optimized a priori based on randomly sampled channels with memory ${L=3}$ and uniformly distributed $\mathsf{snr} \sim \mathcal{U}[0,12]\,\text{dB}$, but \emph{not} specifically on the channel~$\bm{h}_\text{e}$, i.e., it does not require any pilots during the actual transmission.
In the low $\mathsf{snr}$ regime, the EMBP algorithm is close to coherent \ac{MAP} performance with an $\mathsf{snr}$ gap of $1.5\,\text{dB}$ for a target ${\text{BER}=10^{-1}}$. It thereby significantly outperforms the ViterbiNet detector and the \ac{RNN}-based method which have an $\mathsf{snr}$ gap of ${4.5\,\text{dB}}$ and ${8.5\,\text{dB}}$, respectively.
For high $\mathsf{snr}$, the severe \ac{ISI} of the channel leads to a very poor performance of the coherent \ac{BP} detector. The EMBP algorithm performs considerably better but also runs into an error floor. For ${\mathsf{snr} > 7\,\text{dB}}$, it is thereby outperformed by the ViterbiNet detector whose suboptimality mainly originates from an imprecise \ac{CSI} due to the limited amount of training data. The EMBP$^\star$ algorithm has a significantly reduced error floor
and outperforms the ViterbiNet detector in the considered high $\mathsf{snr}$ regime with a ${3\,\text{dB}}$ gain for a target ${\text{BER}=10^{-2}}$. 

\section{Conclusion} \label{sec:conclusion}
We studied the problem of blind joint channel estimation and symbol detection for linear block-fading channels with memory and introduced the EMBP algorithm which interweaves the iterations of \ac{EM} and \ac{BP} on a suitable factor graph. The \ac{VAELE} was found as a suitable method to initialize the EMBP algorithm whose convergence is sensitive to the initialization. We further leveraged techniques of model-based deep learning to develop the refined algorithm EMBP$^\star$ which uses momentum in the updates of \ac{BP} as well as in the \ac{EM} updates to significantly improve the detection performance and reduce the algorithm's complexity with only a few learnable parameters. A complexity analysis and various numerical simulations demonstrate that the proposed EMBP$^\star$ algorithm fulfills the ambition of a fully blind low-complexity symbol detection algorithm that performs comparable to pilot-based or coherent symbol detectors.

\begin{appendix} 
\numberwithin{equation}{subsection}	

\subsection{Derivation of the EM Updates} \label{app:sec:EM}
We consider the optimization problem~\eqref{eq:M-step}. A necessary condition for the optimal solution $\hat{\bm{\theta}}^{(t)}$ is
\begin{align}
    \bm{0} &= \left. \nabla_{\bm{\theta}} \tilde{Q}\mleft(\bm{\theta} \Big\vert \hat{\bm{\theta}}^{(t-1)} \mright) \right|_{\bm{\theta} = \hat{\bm{\theta}}^{(t)}}. \label{app:eq:gradient}
\end{align} %
To rewrite the term
\begin{align}
    \tilde{Q}\mleft(\bm{\theta} \Big\vert \hat{\bm{\theta}}^{(t-1)} \mright) 
    &= \sum\limits_{\bm{c}} P(\bm{c}|\bm{y},\hat{\bm{\theta}}^{(t-1)}) \log p(\bm{y},\bm{c}|\bm{\theta}) \nonumber \\
    &= \sum\limits_{\bm{c}} P(\bm{c}|\bm{y},\hat{\bm{\theta}}^{(t-1)}) \log \mleft( \frac{p(\bm{y}|\bm{c},\bm{\theta})}{M^N} \mright), \nonumber
\end{align}
we use ${P(\bm{c}|\bm{\theta}) = P(\bm{c}) = M^{-N}}$.
The likelihood function $p(\bm{y}|\bm{c},\bm{\theta})$ can be decomposed in the same manner as in~\eqref{eq:expanded_likelihood}. However, note that we must not omit the terms that are independent of $\bm{c}$ in this case, since we are maximizing over $\bm{\theta}$ instead of $\bm{c}$. This leads to
\begin{align}
    \log \mleft( \frac{p(\bm{y}|\bm{c},\bm{\theta})}{M^N} \mright) = &-N \log \mleft( M \pi \sigma^2 \mright) \nonumber \\
    &+ \left( \frac{-\bm{y}^{\rm H}\bm{y} + 2\text{Re}\left\{ \bm{c}^{\rm H} \bm{x} \right\} - \bm{c}^{\rm H} \bm{G} \bm{c} }{\sigma^2} \right) \nonumber \\
    = &-A(\sigma^2) + \frac{-B + C(\bm{c},\bm{h}) - D(\bm{c},\bm{h})}{\sigma^2}, \label{app:eq:likelihood}
\end{align}
where we introduced the terms
\begin{align}
    A(\sigma^2) :=& N \log \mleft( M \pi \sigma^2 \mright), \quad
    B := \sum\limits_{n=1}^{N+L} |y_n|^2, \nonumber \\
    C(\bm{c},\bm{h}) :=& \sum\limits_{n=1}^N \underbrace{2 \text{Re} \left\{ x_n c_n^\star \right\} - G_{n,n} |c_n|^2}_{=: \, C_n(c_n,\bm{h})}, \quad \text{and} \nonumber \\
    D(\bm{c},\bm{h}) :=& \sum\limits_{n=1}^N \sum\limits_{m<n} \underbrace{2 \text{Re} \left\{ G_{m,n} c_m c_n^\star \right\}}_{=: \, D_{n,m}(c_n,c_m,\bm{h})}. \label{app:eq:BCD}
\end{align}
To keep the notation uncluttered, we use the shorter notations ${C(c_n,\bm{h})}$ and ${D(c_n,c_m,\bm{h})}$ for the terms ${C_n(c_n,\bm{h})}$ and ${D_{n,m}(c_n,c_m,\bm{h})}$, respectively.
Note that the terms $\bm{x}$ and $G_{n,m}$ are a function of the channel impulse response $\bm{h}$ which is part of~$\bm{\theta}$. This means that \eqref{app:eq:gradient} is a nonlinear system of equations that can generally not be solved in closed form. However, we can find a closed-form solution for the maximization along each dimension of $\bm{\theta}$, respectively, as suggested by Theorem~\ref{thm:EMSolution}.
Therefore, we solve each row~$i=\ell$ in~\eqref{app:eq:gradient} for one element $\theta_\ell$, respectively, leveraging the formulation of the log-likelihood in \eqref{app:eq:likelihood}. The partial derivative with respect to ${\theta_{L+2}=\sigma^2}$ yields
\begin{align}
    &0 = \left.\dfrac{\partial}{\partial \sigma^2} \tilde{Q}\mleft(\bm{\theta} \Big\vert \hat{\bm{\theta}}^{(t-1)} \mright)\right|_{\sigma^2 = \hat{\sigma}^{2^{(t)}}} \nonumber \\
    \Leftrightarrow \, &0 = \sum\limits_{\bm{c}} P(\bm{c}|\bm{y},\hat{\bm{\theta}}^{(t-1)}) \mleft( \frac{-N}{\hat{\sigma}^{2^{(t)}}} + \frac{B-C(\bm{c},\bm{h})+D(\bm{c},\bm{h})}{\left( \hat{\sigma}^{2^{(t)}} \right)^2} \mright) \nonumber \\
    \Leftrightarrow \, &\hat{\sigma}^{2^{(t)}} = \frac{1}{N} \Bigg[ B - \sum\limits_{n=1}^N \sum\limits_{c_n} P(c_n|\bm{y},\hat{\bm{\theta}}^{(t-1)}) \bigg( C(c_n,\bm{h}) \nonumber \\ 
    &\qquad - \sum\limits_{m<n} \sum\limits_{c_m} P(c_m|\bm{y},\hat{\bm{\theta}}^{(t-1)}) \, D(c_n,c_m,\bm{h}) \bigg) \Bigg], \label{app:eq:sigma2L}
\end{align} %
i.e., we found the closed-form solution~\eqref{app:eq:sigma2L} for the \ac{EM} update of the parameter~$\sigma^2$.

To simplify the derivation of the parameter updates for $\bm{h}$, but without loss of optimality, we split the complex-valued parameters ${h_\ell =: \hRl + \text{j} \hIl}$ into their real and imaginary parts and solve for $\hRl$ and $\hIl$, respectively.
The partial derivative with respect to $\hRl$ yields the condition
\begin{align}
    0 &= \left. \dfrac{\partial}{\partial \hRl} \tilde{Q}\mleft(\bm{\theta} \Big\vert \hat{\bm{\theta}}^{(t-1)} \mright) \right|_{\hRl = \hat{h}_{\text{R},\ell}^{(t)}} \label{app:eq:partial_hR} \\
      &= \sum\limits_{\bm{c}} P(\bm{c}|\bm{y},\hat{\bm{\theta}}^{(t-1)}) \left. \dfrac{\partial}{\partial \hRl} \left( C(\bm{c},\bm{h}) - D(\bm{c},\bm{h}) \right) \right|_{\hRl = \hat{h}_{\text{R},\ell}^{(t)}} . \nonumber
\end{align}
Using the relations
    ${\dfrac{\partial}{\partial \hRl} \text{Re}\{ x_n c_n^\star \} = \text{Re} \{ y_{n+\ell} c_n^\star \}}$
and
    $\dfrac{\partial}{\partial \hRl} G_{n,n} 
    = \dfrac{\partial}{\partial \hRl} \left(\sum\limits_{k=0}^L |h_k|^2 \right)
    = 2 \hRl$, 
we find that
\begin{align}
    \dfrac{\partial}{\partial \hRl} C(\bm{c},\bm{h})
    = & \sum\limits_{n=1}^N \left( 2 \dfrac{\partial}{\partial \hRl} \text{Re} \left\{ x_n c_n^\star \right\} - |c_n|^2 \dfrac{\partial}{\partial \hRl} G_{n,n} \right) \nonumber \\
    = &2 \sum\limits_{n=1}^N \left( \text{Re} \{ y_{n+\ell} c_n^\star \} - |c_n|^2 \hRl \right). \label{app:eq:partial_C}
\end{align}
Substituting \eqref{app:eq:partial_C} into~\eqref{app:eq:partial_hR} yields
\begin{align}
    &\sum\limits_{\bm{c}} P(\bm{c}|\bm{y},\hat{\bm{\theta}}^{(t-1)}) \dfrac{\partial}{\partial \hRl} C(\bm{c},\bm{h}) \label{app:eq:partialC_withSum} \\
    = &2 \sum\limits_{n=1}^N \sum\limits_{c_n} P(c_n|\bm{y},\hat{\bm{\theta}}^{(t-1)}) \left( \text{Re} \{ y_{n+\ell} c_n^\star \} - |c_n|^2 \hRl \right). \nonumber
\end{align}
For the partial derivative of $D(c_n,c_m,\bm{h})$ we obtain
\begin{align}
    &\dfrac{\partial}{\partial \hRl} D(c_n,c_m,\bm{h})
    = 2\dfrac{\partial}{\partial \hRl} \text{Re} \{ G_{m,n} c_m c_n^\star \} \quad \quad \nonumber \\
    =\, &2\dfrac{\partial}{\partial \hRl} \text{Re}\left\{ \sum\limits_{k=0}^{L-(n-m)} h_{k+(n-m)}^\star h_k c_m c_n^\star \right\} \nonumber \\
    =\, \phantom{+} &2\text{Re}\{ c_m c_n^\star \} \left( h_{\text{R},\ell+(n-m)} + h_{\text{R},\ell-(n-m)} \right) \nonumber \\
    + &2\text{Im}\{ c_m c_n^\star \} \left( h_{\text{I},\ell+(n-m)} - h_{\text{I},\ell-(n-m)} \right). \label{app:eq:partial_D}
\end{align}
To rewrite the sums in
\begin{align*}
    \dfrac{\partial}{\partial \hRl} D(\bm{c},\bm{h}) 
    = \sum\limits_{n=1}^N \sum\limits_{m<n} \dfrac{\partial}{\partial \hRl} D(c_n,c_m,\bm{h}),
\end{align*}
we can use the equivalences
\begin{align}
    &\sum\limits_{n=1}^N \sum\limits_{m<n} 2\text{Re}\{ c_m c_n^\star \} \left( h_{\text{R},\ell+(n-m)} + h_{\text{R},\ell-(n-m)} \right) \nonumber \\
    = & \sum\limits_{n=1}^N \sum\limits_{\substack{k=0 \\ k \neq \ell}}^L 2 \text{Re}\{ c_{n-|\ell-k|} c_n^\star \} h_{\text{R},k}, \label{app:eq:sum_simplification_Re}
\end{align}
and
\begin{align}
    &\sum\limits_{n=1}^N \sum\limits_{m<n} 2\text{Im}\{ c_m c_n^\star \} \left( h_{\text{I},\ell+(n-m)} - h_{\text{R},\ell-(n-m)} \right) \nonumber \\
    = & \sum\limits_{n=1}^N \sum\limits_{\substack{k=0 \\ k \neq \ell}}^L 2 \text{Im}\{ c_{n-|\ell-k|} c_n^\star \} \text{sign} \mleft\{ \ell-k \mright\} h_{\text{I},k}. \label{app:eq:sum_simplification_Im}
\end{align} 
Based on~\eqref{app:eq:partial_D} and by using the expressions \eqref{app:eq:sum_simplification_Re} and \eqref{app:eq:sum_simplification_Im}, we find for the second term in~\eqref{app:eq:partial_hR}:
\begin{align}
    &\sum\limits_{\bm{c}} P(\bm{c}|\bm{y},\hat{\bm{\theta}}^{(t-1)}) \dfrac{\partial}{\partial \hRl} D(\bm{c},\bm{h}) \nonumber \\
    = 2 &\sum\limits_{n=1}^N \sum\limits_{c_n} P(c_n|\bm{y},\hat{\bm{\theta}}^{(t-1)}) \sum\limits_{\substack{k=0 \\ k \neq \ell}}^L \sum\limits_{c_{n-|\ell-k|}} \hspace{-6pt} P(c_{n-|\ell-k|}|\bm{y},\hat{\bm{\theta}}^{(t-1)}) \nonumber \\ 
    &\left( \text{Re}\{ c_{n-|\ell-k|} c_n^\star \} h_{\text{R},k} + \text{Im}\{ c_{n-|\ell-k|} c_n^\star \} \text{sign} \mleft\{ \ell-k \mright\} h_{\text{I},k} \right). \label{app:eq:partialD_withSum}
\end{align}
Applying \eqref{app:eq:partialC_withSum} and \eqref{app:eq:partialD_withSum} in~\eqref{app:eq:partial_hR} and solving for $\hRl$ finally yields the update equation for $\hat{h}_{\text{R},\ell}^{(t)}$ given in~\eqref{app:eq:update_hRl}.
In a similar fashion, we can solve
    $0 = {\dfrac{\partial}{\partial \hIl} \tilde{Q}\mleft(\bm{\theta} \Big\vert \hat{\bm{\theta}}^{(t-1)} \mright)}$ %
for $\hIl$ which leads to the update equation~\eqref{app:eq:update_hIl} for $\hat{h}_{\text{I},\ell}^{(t)}$.
\begin{figure*}
\begin{align}
&\begin{aligned}
     \hat{h}_{\text{R},\ell}^{(t)} 
     =  \Bigg( &\sum\limits_{n=1}^N \sum\limits_{c_n} P(c_n|\bm{y},\hat{\bm{\theta}}^{(t-1)})  \text{Re} \{ y_{n+\ell} c_n^\star \} - \sum\limits_{\substack{k=0 \\ k \neq \ell}}^L \sum\limits_{c_{n-|\ell-k|}} P(c_{n-|\ell-k|}|\bm{y},\hat{\bm{\theta}}^{(t-1)}) \\
     &\bigg( \text{Re}\{ c_{n-|\ell-k|} c_n^\star \} h_{\text{R},k}
     + \text{Im}\{ c_{n-|\ell-k|} c_n^\star \} \text{sign} \mleft\{ \ell-k \mright\} h_{\text{I},k} \bigg) \Bigg) 
     \cdot \left( \sum\limits_{n=1}^N \sum\limits_{c_n} P(c_n|\bm{y},\hat{\bm{\theta}}^{(t-1)}) |c_n|^2 \right)^{-1}
\end{aligned} \label{app:eq:update_hRl} \\
&\begin{aligned}
     \hat{h}_{\text{I},\ell}^{(t)} 
     = \Bigg( &\sum\limits_{n=1}^N \sum\limits_{c_n} P(c_n|\bm{y},\hat{\bm{\theta}}^{(t-1)})  \text{Im} \{ y_{n+\ell} c_n^\star \} - \sum\limits_{\substack{k=0 \\ k \neq \ell}}^L \sum\limits_{c_{n-|\ell-k|}} P(c_{n-|\ell-k|}|\bm{y},\hat{\bm{\theta}}^{(t-1)}) \\
     &\bigg( \text{Re}\{ c_{n-|\ell-k|} c_n^\star \} h_{\text{I},k}
     - \text{Im}\{ c_{n-|\ell-k|} c_n^\star \} \text{sign} \mleft\{ \ell-k \mright\} h_{\text{R},k} \bigg) \Bigg) 
     \cdot \left( \sum\limits_{n=1}^N \sum\limits_{c_n} P(c_n|\bm{y},\hat{\bm{\theta}}^{(t-1)}) |c_n|^2 \right)^{-1} 
\end{aligned} \label{app:eq:update_hIl} %
\end{align}
\hrule
\end{figure*}
By studying the two equations~\eqref{app:eq:update_hRl} and~\eqref{app:eq:update_hIl}, we can observe that the update of $\hat{h}_{\text{R},\ell}^{(t)}$ is independent of $\hat{h}_{\text{I},\ell}^{(t-1)}$ and that, vice versa, the update of $\hat{h}_{\text{I},\ell}^{(t)}$ does not depend on $\hat{h}_{\text{R},\ell}^{(t-1)}$. Therefore, we can update $\hat{h}_{\text{R},\ell}^{(t)}$ and $\hat{h}_{\text{I},\ell}^{(t)}$ in parallel without loss of optimality, i.e., 
    ${\hat{h}_\ell = \hat{h}_{\text{R},\ell}^{(t)} + \text{j} \hat{h}_{\text{I},\ell}^{(t)}}$, %
thus proving the theorem. 
\qed

\subsection{Complexity} \label{app:complexity}
We analyze the complexity of the proposed EMBP algorithm by counting the number of real-valued additions~(ADD), multiplications~(MULT) and ${\log \Sigma \exp}$-operations that are required in the \Estep~(1 \ac{BP} iteration) and in the \Mstep\, to update the parameter estimates $\hat{\sigma}^2$ and $\hat{h}_\ell$. Complex additions are counted as 2 ADD operations and complex multiplications are counted as 4 MULT and 2 ADD operations. We neglect interim results that are data-independent, e.g., $|c_n|^2$, as they can be precomputed and stored. Also note that, depending on the parameter update schedule of the EMBP algorithm, some data-dependent interim results can be reused. For simplicity, however, we consider them to be recomputed each time. The results are provided in Table~\ref{tab:complexity}.\\
The EMBP$^\star$ algorithm requires some extra operations for the convex combination in the momentum updates. For the momentum updates of the \ac{BP} messages, ${N(2L+1)}$ real-valued messages of dimension $M$ are scaled and added to a scaled version of the old messages, which results in ${2NM(2L+1)}$~MULT and ${NM(2L+1)}$~ADD operations. For the momentum-based parameter update of one complex-valued element in $\hat{\bm{\theta}}$, it requires 4~MULT and 2~ADD operations.

We further provide some remarks about the complexity of the pilot-based and decision-directed MAP detection schemes that are considered in Sec.~\ref{sec:experiments} for comparison.
For the pilot-based MAP detector with $N_\text{p}$ pilots, the complexity of the MAP detector (see Table~\ref{tab:complexity}) is supplemented by the complexity of the pilot-based ML estimator, which can be implemented by an ${(L+1)\times N_\text{p}}$ complex-valued matrix-vector multiplication. The DD-MAP scheme builds upon the previously mentioned pilot-based MAP detector. Additionally, it requires the ML estimation based on the hard symbol decision of the first MAP detection run. This can be implemented by an ${(L+1)\times N}$ complex-valued matrix-vector multiplication and an algorithm to solve an $(L+1)\times(L+1)$ Toeplitz system, e.g., the Levinson-Durbin recursion. Furthermore, a second run of the MAP detector is required.

\subsection{FEC-BER Performance}\label{app:fec}
We consider an \ac{FEC} scheme applied to the \ac{BPSK} transmission scenario with block length ${N=100}$ and memory ${L=5}$, as described in Sec.~\ref{sec:experiments} with the results shown in Fig.~\ref{fig:MSE_vs_SNR_BPSK_L5} and Fig.~\ref{fig:BER_vs_SNR_BPSK_L5}. 
For the non-blind baseline algorithms with $20\%$ pilots we apply a \ac{LDPC} code of rate~$0.8$, and for the blind EMBP$^\star$ algorithm, we use an \ac{LDPC} code of rate~$0.64$. Thereby, all compared schemes effectively transmit $64$ information bits. For channel decoding, we use a normalized min-sum decoder with a maximum number of 20 layered BP iterations. Fig.~\ref{app:fig:fec_ber_qpsk_L5} compares the \ac{BER} and \ac{FEC}-\ac{BER} performance of the EMBP$^\star$ algorithm with the pilot-based schemes. With \ac{FEC}, the EMBP$^\star$ algorithm consistently outperforms the pilot-based \ac{MAP} detector with a $2\,\text{dB}$ \ac{SNR} gain and the pilot-based DD-MAP baseline with a gain of $1\,\text{dB}$.
\begin{figure}[tb]
  \centering
      \begin{tikzpicture}
    \begin{axis}[
    width=\linewidth, %
    height=0.75\linewidth,
    align = left,
    grid=major, %
    grid style={gray!30}, %
    xlabel= $\mathsf{snr}$ (dB),
    ylabel= BER,
	  scaled y ticks=false,
    ymode = log,
    ymin = 0.00001,
    ymax = 0.5,
    xmin = -4,
    xmax = 12,
    enlarge x limits=false,
    enlarge y limits=false,
    line width=1pt,
	  legend style={font=\footnotesize, cells={align=left}, anchor=south west, at={(0.015,0.015)}},
    legend cell align={left},
	  smooth,
    ]
    \addlegendentry{MAP, $\hat{\bm{h}}_\text{ML}$, \\$20\%$ pilots}
     \addplot[color=black!60!white, line width=1pt, dashdotted] table[x={Es/N0 (dB)}, y={BER MAP pilots}, col sep=comma] {numerical_results/decdir_BPSK_L5_20pilots.csv};
     
    \addlegendentry{MAP, $\hat{\bm{h}}_\text{DD-MAP}$, \\$20\%$ pilots}
    \addplot[color=decdircolor!60!white, line width=1pt, dashdotted] table[x={Es/N0 (dB)}, y={BER MAP decdirMAP}, col sep=comma] {numerical_results/decdir_BPSK_L5_20pilots.csv};
   
    \addplot[color=EMNBPcolor, line width=1pt] table[x={EsN0 dB}, y={EMNBP(VAE) BER}, col sep=comma] {numerical_results/BER+MSE_over_EsN0_BPSK_L5.csv};
    \addlegendentry{EMBP$^\star$, $\beta_\text{BP}^\star$}
    \addlegendentry{with FEC}
    \addplot[draw=none,color=black, line width=1pt, solid, mark=*, mark options={solid}] table[x={Es/N0 (dB)}, y={BER MAP pilots}, col sep=comma] {numerical_results/BER_coded_BPSK.csv};
    
     \addplot[color=black!60!white, line width=1pt, dashdotted, mark=*, mark options={solid}] table[x={Es/N0 (dB)}, y={BER MAP pilots}, col sep=comma] {numerical_results/BER_coded_BPSK.csv};
    \addplot[color=decdircolor!60!white, line width=1pt, dashdotted, mark=*, mark options={solid}] table[x={Es/N0 (dB)}, y={BER MAP decdirMAP coded}, col sep=comma] {numerical_results/BER_coded_BPSK.csv};
    \addplot[color=EMNBPcolor, line width=1pt, mark=*, mark options={solid}] table[x={Es/N0 (dB)}, y={BER EMNBP coded}, col sep=comma] {numerical_results/BER_coded_BPSK.csv};
    
  \end{axis}
\end{tikzpicture}
    \caption{BER and FEC-BER over \textsf{snr} for a BPSK transmission over $10^7$ random channels with ${L=5}$. The FEC-BER results are based on an LDPC code of rate $0.8$ for the pilot-based schemes and a code rate of $0.64$ for the EMBP$^\star$ algorithm, such that all compared schemes effectively transmit $64$ information bits.}
  \label{app:fig:fec_ber_qpsk_L5}
\end{figure}

\subsection{Channels with Non-uniform \ac{pdp}}\label{app:exp_channels}
We evaluate the estimation and detection performance of the EMBP algorithm on $10^7$ random channels with a non-uniformly distributed \ac{pdp}. In contrast to the channel model defined in Sec.~\ref{subsec:system_model}, the channel taps $h_\ell$, ${\ell=0,\ldots,L}$ are independently sampled from a complex-circular Gaussian distribution with exponentially decaying variance~$\mathcal{N}(0,\exp(-\ell))$. For better comparison, the power of the sampled channel impulse responses is normalized as before. Fig.~\ref{app:fig:mse_L5exp} plots the squared estimation error over the ${\mathsf{snr}}$ and Fig.~\ref{app:fig:ber_L5exp} shows the respective \ac{BER} performance. \\
We additionally simulate the transmission of ${N=100}$ \ac{QPSK} symbols per transmission block on the same batch of $10^7$ channels. The evaluation results for the squared estimation error and the \ac{BER} over ${\mathsf{snr}}$ are shown in Fig.~\ref{app:fig:mse_qpsk_L5exp} and Fig.~\ref{app:fig:ber_qpsk_L5exp}, respectively.
\begin{figure}[tb]
\centering
   \begin{tikzpicture}
    \pgfplotsset{
        legend image VAELE/.style={
            legend image code/.code={%
            \draw[solid, color=VAEcolor] (0cm,0.05cm) -- (0.3cm,0.05cm);
            \draw[dashed, color=VAEcolor] (0cm, -0.05cm) -- (0.3cm, -0.05cm);
            }
        },
    }
        \pgfplotsset{
        legend image EMBPb/.style={
            legend image code/.code={%
            \draw[solid, color=EMBPDiraccolor] (0cm,0.05cm) -- (0.3cm,0.05cm);
            \draw[dashed, color=EMBPDiraccolor] (0cm, -0.05cm) -- (0.3cm, -0.05cm);
            }
        },
    }
        \pgfplotsset{
        legend image EMBPc/.style={
            legend image code/.code={%
            \draw[solid,color=EMBPcolor] (0cm,0.05cm) -- (0.3cm,0.05cm);
            \draw[dashed,color=EMBPcolor] (0cm, -0.05cm) -- (0.3cm, -0.05cm);
            }
        },
    }
    \pgfplotsset{
        legend image EMNBPc/.style={
            legend image code/.code={%
            \draw[solid,color=EMNBPcolor] (0cm,0.05cm) -- (0.3cm,0.05cm);
            \draw[dashed,color=EMNBPcolor] (0cm, -0.05cm) -- (0.3cm, -0.05cm);
            }
        },
    }
    \pgfplotsset{
        legend image ML/.style={
            legend image code/.code={%
            \draw[color=black!100!white] (0cm,0.05cm) -- (0.3cm,0.05cm);
            \draw[color=black!60!white] (0cm, -0.05cm) -- (0.3cm, -0.05cm);
            }
        },
    }
    \pgfplotsset{
        legend image decdir/.style={
            legend image code/.code={%
            \draw[color=decdircolor!100!white] (0cm,0.05cm) -- (0.3cm,0.05cm);
            \draw[color=decdircolor!60!white] (0cm, -0.05cm) -- (0.3cm, -0.05cm);
            }
        },
    }
    \pgfplotsset{
legend image code/.code={
\draw[mark repeat=2,mark phase=2]
plot coordinates {
(0cm,0cm)
(0.15cm,0cm)        %
(0.3cm,0cm)         %
};%
}
}
    \begin{axis}[
    width=\linewidth, %
    height=0.7\linewidth,
    align = left,
    grid=major, %
    grid style={gray!30}, %
    xlabel= $\mathsf{snr}$ (dB),
    ylabel= ${\lVert \hat{\boldsymbol{h}} - \boldsymbol{h} \rVert^2}$,
	  scaled y ticks=false,
    ymode = log,
    ymin = 0.002,
    ymax = 1.0,
    xmin = -4,
    xmax = 12,
    enlarge x limits=false,
    enlarge y limits=false,
    line width=1pt,
	  legend style={font=\footnotesize, cells={align=left}, anchor=south west, at={(0.015,0.015)}},
    legend cell align={left},
    ]
    \addlegendimage{legend image VAELE}
    \addlegendentry{VAE-LE}
    \addlegendimage{legend image EMBPb}
   \addlegendentry{EMBP, init.~(b)}
    \addlegendimage{legend image EMBPc}
    \addlegendentry{EMBP, init.~(c)}
    \addlegendimage{legend image ML}
    \addlegendentry{ML, $10/20\%$ pilots}
    \addlegendimage{legend image decdir}
    \addlegendentry{DD-MAP, $10/20\%$ pilots}

    \addplot[color=black!100!white, line width=1pt] table[x={Es/N0 (dB)}, y={MSE h pilots}, col sep=comma] {numerical_results/decdir_BPSK_L5exp_10pilots.csv};
    \node at (axis cs:8,0.5) {\footnotesize ${10\%}$};
    \addplot[color=black!60!white, line width=1pt] table[x={Es/N0 (dB)}, y={MSE h pilots}, col sep=comma] {numerical_results/decdir_BPSK_L5exp_20pilots.csv};
    \node at (axis cs:10,0.07) {\footnotesize \color{black!60!white} ${20\%}$};

    \addplot[color=decdircolor!100!white, line width=1pt] table[x={Es/N0 (dB)}, y={MSE h decdirMAP}, col sep=comma] {numerical_results/decdir_BPSK_L5exp_10pilots.csv};
    \node at (axis cs:8,0.04) {\footnotesize \color{decdircolor!100!white} ${10\%}$};
    \addplot[color=decdircolor!60!white, line width=1pt] table[x={Es/N0 (dB)}, y={MSE h decdirMAP}, col sep=comma] {numerical_results/decdir_BPSK_L5exp_20pilots.csv};
    \node at (axis cs:3,0.12) {\footnotesize \color{decdircolor!60!white} ${20\%}$};
    
	\addplot[color=VAEcolor, line width=1pt] table[x={EsN0 dB}, y={VAE meanSE}, col sep=comma] {numerical_results/BER+MSE_over_EsN0_BPSK_L5exp.csv};
    \addplot[color=VAEcolor, line width=1pt, dashed, mark options={solid}] table[x={EsN0 dB}, y={VAE medianSE}, col sep=comma] {numerical_results/BER+MSE_over_EsN0_BPSK_L5exp.csv};
    \addplot[color=EMBPDiraccolor, line width=1pt] table[x={EsN0 dB}, y={EMBP(Dirac) meanSE}, col sep=comma] {numerical_results/BER+MSE_over_EsN0_BPSK_L5exp.csv};
    \addplot[color=EMBPDiraccolor, line width=1pt, dashed] table[x={EsN0 dB}, y={EMBP(Dirac) medianSE}, col sep=comma] {numerical_results/BER+MSE_over_EsN0_BPSK_L5exp.csv};
    \addplot[color=EMBPcolor, line width=1pt] table[x={EsN0 dB}, y={EMBP(VAE) meanSE}, col sep=comma] {numerical_results/BER+MSE_over_EsN0_BPSK_L5exp.csv};
    \addplot[color=EMBPcolor, line width=1pt, dashed] table[x={EsN0 dB}, y={EMBP(VAE) medianSE}, col sep=comma] {numerical_results/BER+MSE_over_EsN0_BPSK_L5exp.csv};
  \end{axis}
\end{tikzpicture}
    \caption{MSE over $\mathsf{snr}$ for various algorithms (solid line: mean, dashed line: median). Evaluation for a BPSK transmission over $10^7$ random channels with ${L=5}$ and exponentially decaying PDP.}
  \label{app:fig:mse_L5exp}
\end{figure}%
\begin{figure}
  \centering
      \begin{tikzpicture}
      \pgfplotsset{
        legend image solid/.style={
            legend image code/.code={%
            \draw[color=black!100!white] (0cm,0.05cm) -- (0.3cm,0.05cm);
            \draw[color=black!60!white] (0cm, -0.05cm) -- (0.3cm, -0.05cm);
            }
        },
    }
    \pgfplotsset{
        legend image dotted/.style={
            legend image code/.code={%
            \draw[dashdotted, color=black!100!white] (0cm,0.05cm) -- (0.3cm,0.05cm);
            \draw[dashdotted, color=black!60!white] (0cm, -0.05cm) -- (0.3cm, -0.05cm);
            }
        },
    }
    \pgfplotsset{
        legend image decdir/.style={
            legend image code/.code={%
            \draw[dashdotted, color=decdircolor!100!white] (0cm,0.05cm) -- (0.3cm,0.05cm);
            \draw[dashdotted, color=decdircolor!60!white] (0cm, -0.05cm) -- (0.3cm, -0.05cm);
            }
        },
    }
    \pgfplotsset{
legend image code/.code={
\draw[mark repeat=2,mark phase=2]
plot coordinates {
(0cm,0cm)
(0.15cm,0cm)        %
(0.3cm,0cm)         %
};%
}
}
    \begin{axis}[
    width=\linewidth, %
    height=0.8\linewidth,
    align = left,
    grid=major, %
    grid style={gray!30}, %
    xlabel= $\mathsf{snr}$ (dB),
    ylabel= BER,
	  scaled y ticks=false,
    ymode = log,
    ymin = 0.0003,
    ymax = 0.5,
    xmin = -4,
    xmax = 12,
    enlarge x limits=false,
    enlarge y limits=false,
    line width=1pt,
	  legend style={font=\footnotesize, cells={align=left}, anchor=south west, at={(0.015,0.015)}},
    legend cell align={left},
	  smooth,
    ]
    \addplot[color=VAEcolor, line width=1pt] table[x={EsN0 dB}, y={VAE BER}, col sep=comma] {numerical_results/BER+MSE_over_EsN0_BPSK_L5exp.csv};
    \addlegendentry{VAE-LE}
    \addplot[color=EMBPcolor, line width=1pt] table[x={EsN0 dB}, y={EMBP(VAE) BER}, col sep=comma] {numerical_results/BER+MSE_over_EsN0_BPSK_L5exp.csv};
    \addlegendentry{EMBP}
    \addplot[color=EMNBPcolor, line width=1pt] table[x={EsN0 dB}, y={EMNBP(VAE) BER}, col sep=comma] {numerical_results/BER+MSE_over_EsN0_BPSK_L5exp.csv};
    \addlegendentry{EMBP$^\star$, $\beta_\text{BP}^\star$}
    \addplot[color=BPcolor, line width=1pt, densely dotted] table[x={EsN0 dB}, y={BP(genie) BER}, col sep=comma] {numerical_results/BER+MSE_over_EsN0_BPSK_L5exp.csv};
    \addlegendentry{BP, $\bm{h}$ (coherent)}
    \addplot[color=black!30!white, line width=1pt, densely dotted] table[x={EsN0 dB}, y={MAP BER}, col sep=comma] {numerical_results/MAP_BER_over_EsN0_BPSK_L5exp.csv};
    \addlegendentry{MAP, $\bm{h}$ (coherent)}
    \addlegendimage{legend image dotted}
    \addlegendentry{MAP, $\hat{\bm{h}}_\text{ML}$, \\ $10/20\%$ pilots}
    \addlegendimage{legend image decdir}
    \addlegendentry{MAP, $\hat{\bm{h}}_\text{DD-MAP}$, \\ $10/20\%$ pilots}
    \addplot[color=black!100!white, line width=1pt, dashdotted] table[x={Es/N0 (dB)}, y={BER MAP pilots}, col sep=comma] {numerical_results/decdir_BPSK_L5exp_10pilots.csv};

    \addplot[color=black!60!white, line width=1pt, dashdotted] table[x={Es/N0 (dB)}, y={BER MAP pilots}, col sep=comma] {numerical_results/decdir_BPSK_L5exp_20pilots.csv};
    \node at (axis cs:10,0.03) {\footnotesize \color{black!100!white} ${10\%}$};
    \node at (axis cs:10.5,0.0005) {\footnotesize \color{black!60!white} ${20\%}$};
     
    \addplot[color=decdircolor!100!white, line width=1pt, dashdotted] table[x={Es/N0 (dB)}, y={BER MAP decdirMAP}, col sep=comma] {numerical_results/decdir_BPSK_L5exp_10pilots.csv};
    \addplot[color=decdircolor!60!white, line width=1pt, dashdotted] table[x={Es/N0 (dB)}, y={BER MAP decdirMAP}, col sep=comma] {numerical_results/decdir_BPSK_L5exp_20pilots.csv};
    \node at (axis cs:7.4,0.03) {\footnotesize \color{decdircolor!100!white} ${10\%}$};
    \node at (axis cs:8.8,0.0005) {\footnotesize \color{decdircolor!60!white} ${20\%}$};
  \end{axis}
\end{tikzpicture}
    \caption{BER over \textsf{snr} for a BPSK transmission over $10^7$ random channels with ${L=5}$ and exponentially decaying PDP.}
\label{app:fig:ber_L5exp}
\end{figure}
\begin{figure}[h]
  \centering
  \begin{tikzpicture}
    \pgfplotsset{
        legend image VAELE/.style={
            legend image code/.code={%
            \draw[solid, color=VAEcolor] (0cm,0.05cm) -- (0.3cm,0.05cm);
            \draw[dashed, color=VAEcolor] (0cm, -0.05cm) -- (0.3cm, -0.05cm);
            }
        },
    }
        \pgfplotsset{
        legend image EMBPb/.style={
            legend image code/.code={%
            \draw[solid, color=EMBPDiraccolor] (0cm,0.05cm) -- (0.3cm,0.05cm);
            \draw[dashed, color=EMBPDiraccolor] (0cm, -0.05cm) -- (0.3cm, -0.05cm);
            }
        },
    }
        \pgfplotsset{
        legend image EMBPc/.style={
            legend image code/.code={%
            \draw[solid,color=EMBPcolor] (0cm,0.05cm) -- (0.3cm,0.05cm);
            \draw[dashed,color=EMBPcolor] (0cm, -0.05cm) -- (0.3cm, -0.05cm);
            }
        },
    }
    \pgfplotsset{
        legend image EMNBPc/.style={
            legend image code/.code={%
            \draw[solid,color=EMNBPcolor] (0cm,0.05cm) -- (0.3cm,0.05cm);
            \draw[dashed,color=EMNBPcolor] (0cm, -0.05cm) -- (0.3cm, -0.05cm);
            }
        },
    }
    \pgfplotsset{
        legend image ML/.style={
            legend image code/.code={%
            \draw[color=black!100!white] (0cm,0.05cm) -- (0.3cm,0.05cm);
            \draw[color=black!60!white] (0cm, -0.05cm) -- (0.3cm, -0.05cm);
            }
        },
    }
    \pgfplotsset{
        legend image decdir/.style={
            legend image code/.code={%
            \draw[color=decdircolor!100!white] (0cm,0.05cm) -- (0.3cm,0.05cm);
            \draw[color=decdircolor!60!white] (0cm, -0.05cm) -- (0.3cm, -0.05cm);
            }
        },
    }
    \pgfplotsset{
legend image code/.code={
\draw[mark repeat=2,mark phase=2]
plot coordinates {
(0cm,0cm)
(0.15cm,0cm)        %
(0.3cm,0cm)         %
};%
}
}
    \begin{axis}[
    width=\linewidth, %
    height=0.67\linewidth,
    align = left,
    grid=major, %
    grid style={gray!30}, %
    xlabel= $\mathsf{snr}$ (dB),
    ylabel= ${\lVert \hat{\boldsymbol{h}} - \boldsymbol{h} \rVert^2}$,
	  scaled y ticks=false,
    ymode = log,
    ymin = 0.002,
    ymax = 1.0,
    xmin = -4,
    xmax = 12,
    enlarge x limits=false,
    enlarge y limits=false,
    line width=1pt,
	  legend style={font=\footnotesize, cells={align=left}, anchor=south west, at={(0.015,0.015)}},
    legend cell align={left},
    ]
    \addlegendimage{legend image VAELE}
    \addlegendentry{VAE-LE}
    \addlegendimage{legend image EMBPb}
   \addlegendentry{EMBP, init.~(b)}
    \addlegendimage{legend image EMBPc}
    \addlegendentry{EMBP, init.~(c)}
    \addlegendimage{legend image ML}
    \addlegendentry{ML, $10/20\%$ pilots}
    \addlegendimage{legend image decdir}
    \addlegendentry{DD-MAP, $10/20\%$ pilots}

    \addplot[color=black!100!white, line width=1pt] table[x={Es/N0 (dB)}, y={MSE h pilots}, col sep=comma] {numerical_results/decdir_QPSK_L5exp_10pilots.csv};
    \node at (axis cs:6.5,0.6) {\footnotesize ${10\%}$};
    \addplot[color=black!60!white, line width=1pt] table[x={Es/N0 (dB)}, y={MSE h pilots}, col sep=comma] {numerical_results/decdir_QPSK_L5exp_20pilots.csv};
    \node at (axis cs:4,0.27) {\footnotesize \color{black!60!white} ${20\%}$};

    \addplot[color=decdircolor!100!white, line width=1pt] table[x={Es/N0 (dB)}, y={MSE h decdirMAP}, col sep=comma] {numerical_results/decdir_QPSK_L5exp_10pilots.csv};
    \node at (axis cs:9,0.15) {\footnotesize \color{decdircolor!100!white} ${10\%}$};
    \addplot[color=decdircolor!60!white, line width=1pt] table[x={Es/N0 (dB)}, y={MSE h decdirMAP}, col sep=comma] {numerical_results/decdir_QPSK_L5exp_20pilots.csv};
    \node at (axis cs:11,0.004) {\footnotesize \color{decdircolor!60!white} ${20\%}$};
    
	\addplot[color=VAEcolor, line width=1pt] table[x={EsN0 dB}, y={VAE meanSE}, col sep=comma] {numerical_results/BER+MSE_over_EsN0_QPSK_L5exp.csv};
    \addplot[color=VAEcolor, line width=1pt, dashed, mark options={solid}] table[x={EsN0 dB}, y={VAE medianSE}, col sep=comma] {numerical_results/BER+MSE_over_EsN0_QPSK_L5exp.csv};
    \addplot[color=EMBPDiraccolor, line width=1pt] table[x={EsN0 dB}, y={EMBP(Dirac) meanSE}, col sep=comma] {numerical_results/BER+MSE_over_EsN0_QPSK_L5exp.csv};
    \addplot[color=EMBPDiraccolor, line width=1pt, dashed] table[x={EsN0 dB}, y={EMBP(Dirac) medianSE}, col sep=comma] {numerical_results/BER+MSE_over_EsN0_QPSK_L5exp.csv};
    \addplot[color=EMBPcolor, line width=1pt] table[x={EsN0 dB}, y={EMBP(VAE) meanSE}, col sep=comma] {numerical_results/BER+MSE_over_EsN0_QPSK_L5exp.csv};
    \addplot[color=EMBPcolor, line width=1pt, dashed] table[x={EsN0 dB}, y={EMBP(VAE) medianSE}, col sep=comma] {numerical_results/BER+MSE_over_EsN0_QPSK_L5exp.csv};
  \end{axis}
\end{tikzpicture}
\caption{MSE over $\mathsf{snr}$ for various algorithms (solid line: mean, dashed line: median). Evaluation for a QPSK transmission over $10^7$ random channels with ${L=5}$ and exponentially decaying PDP.}
\label{app:fig:mse_qpsk_L5exp}
\end{figure}
\begin{figure}[tb]
  \centering
      \begin{tikzpicture}
      \pgfplotsset{
        legend image solid/.style={
            legend image code/.code={%
            \draw[color=black!100!white] (0cm,0.05cm) -- (0.3cm,0.05cm);
            \draw[color=black!60!white] (0cm, -0.05cm) -- (0.3cm, -0.05cm);
            }
        },
    }
    \pgfplotsset{
        legend image dotted/.style={
            legend image code/.code={%
            \draw[dashdotted, color=black!100!white] (0cm,0.05cm) -- (0.3cm,0.05cm);
            \draw[dashdotted, color=black!60!white] (0cm, -0.05cm) -- (0.3cm, -0.05cm);
            }
        },
    }
    \pgfplotsset{
        legend image decdir/.style={
            legend image code/.code={%
            \draw[dashdotted, color=decdircolor!100!white] (0cm,0.05cm) -- (0.3cm,0.05cm);
            \draw[dashdotted, color=decdircolor!60!white] (0cm, -0.05cm) -- (0.3cm, -0.05cm);
            }
        },
    }
    \pgfplotsset{
legend image code/.code={
\draw[mark repeat=2,mark phase=2]
plot coordinates {
(0cm,0cm)
(0.15cm,0cm)        %
(0.3cm,0cm)         %
};%
}
}
    \begin{axis}[
    width=\linewidth, %
    height=0.7\linewidth,
    align = left,
    grid=major, %
    grid style={gray!30}, %
    xlabel= $\mathsf{snr}$ (dB),
    ylabel= BER,
	  scaled y ticks=false,
    ymode = log,
    ymin = 0.0003,
    ymax = 0.5,
    xmin = -4,
    xmax = 12,
    enlarge x limits=false,
    enlarge y limits=false,
    line width=1pt,
	  legend style={font=\footnotesize, cells={align=left}, anchor=south west, at={(0.015,0.015)}},
    legend cell align={left},
	  smooth,
    ]
    \addplot[color=EMNBPcolor, line width=1pt] table[x={EsN0 dB}, y={EMNBP(VAE) BER}, col sep=comma] {numerical_results/BER+MSE_over_EsN0_QPSK_L5exp.csv};
    \addlegendentry{EMBP$^\star$, $\beta_\text{BP}^\star$}
    \addplot[color=BPcolor, line width=1pt, densely dotted] table[x={EsN0 dB}, y={BP(genie) BER}, col sep=comma] {numerical_results/BER+MSE_over_EsN0_QPSK_L5exp.csv};
    \addlegendentry{BP, $\bm{h}$ (coherent)}
    \addplot[color=black!30!white, line width=1pt, densely dotted] table[x={EsN0 dB}, y={MAP BER}, col sep=comma] {numerical_results/MAP_BER_over_EsN0_QPSK_L5exp.csv};
    \addlegendentry{MAP, $\bm{h}$ (coherent)}
    \addlegendimage{legend image dotted}
    \addlegendentry{MAP, $\hat{\bm{h}}_\text{ML}$, \\ $10/20\%$ pilots}
    \addlegendimage{legend image decdir}
    \addlegendentry{MAP, $\hat{\bm{h}}_\text{DD-MAP}$, \\ $10/20\%$ pilots}
    \addplot[color=black!100!white, line width=1pt, dashdotted] table[x={Es/N0 (dB)}, y={BER MAP pilots}, col sep=comma] {numerical_results/decdir_QPSK_L5exp_10pilots.csv};

    \addplot[color=black!60!white, line width=1pt, dashdotted] table[x={Es/N0 (dB)}, y={BER MAP pilots}, col sep=comma] {numerical_results/decdir_QPSK_L5exp_20pilots.csv};
    \node at (axis cs:8,0.15) {\footnotesize \color{black!100!white} ${10\%}$};
    \node at (axis cs:11.4,0.004) {\tiny \color{black!60!white} ${20\%}$};
     
    \addplot[color=decdircolor!100!white, line width=1pt, dashdotted] table[x={Es/N0 (dB)}, y={BER MAP decdirMAP}, col sep=comma] {numerical_results/decdir_QPSK_L5exp_10pilots.csv};
    \addplot[color=decdircolor!60!white, line width=1pt, dashdotted] table[x={Es/N0 (dB)}, y={BER MAP decdirMAP}, col sep=comma] {numerical_results/decdir_QPSK_L5exp_20pilots.csv};
    \node at (axis cs:11.1,0.02) {\footnotesize \color{decdircolor!100!white} ${10\%}$};
    \node at (axis cs:11.3,0.001) {\tiny \color{decdircolor!60!white} ${20\%}$};
  \end{axis}
\end{tikzpicture}
    \caption{BER over \textsf{snr} for a QPSK transmission over $10^7$ random channels with ${L=5}$ and exponentially decaying PDP.}
  \label{app:fig:ber_qpsk_L5exp}
\end{figure}
\end{appendix}

\end{document}